\documentclass[preprint2]{aastex}

\newcommand \dpa {\Delta\phi_{\rm PA}}
\newcommand \nin {\noindent}

\newcommand{\rovc}{r_{\rm ovc}}

\newcommand{\rns}{R_{\rm ns}}
\newcommand{\rlc}{R_{\rm lc}}
\newcommand{\rmax}{r_{\rm max}}

\newcommand \vrd {\vec r}
\newcommand \rvers {\hat r}
\newcommand \kp {\hat k^\prime}
\newcommand \ka {\hat k}
\newcommand \bfl {\vec B}
\newcommand \n {\hat n}
\newcommand \phecl {\phi_{\rm ecl}}
\newcommand \phem {\phi_{\rm em}}
\newcommand \phdip {\phi_m^{\rm srf}}
\newcommand \dshap {\Delta\phi_{\rm Sh}}
\newcommand \al {\alpha}
\newcommand \ze {\zeta}
\newcommand \thb {\theta_b}
\newcommand \thm {\theta_m}
\newcommand \thab {\theta_{\rm ab}}
\newcommand \vmu {\vec\mu}
\newcommand \s {s}
\newcommand \btcor {\beta_c}

\newcommand \vvcor {\vec v_c}
\newcommand \vomega {\vec\Omega}
\newcommand \gamc {\gamma_c}
\newcommand \reff {R_{\rm eff}}
\newcommand \recl {r_{\rm ecl}}
\newcommand \hecl {h_{\rm ecl}}
\newcommand \lt {{\rm LT}}
\newcommand \vA {\vec A^\prime}
\newcommand \rhodip {\rho_m}
\newcommand \thpc {\theta_{\rm pc}}
\newcommand \thpcret {\theta_{\rm pc}^{\rm ret}}
\newcommand \thsurf {\theta_m^{\rm srf}}
\newcommand \rmin {r_{\rm min}}
\newcommand \zmax {z_{\rm max}}
\newcommand \rhomax {\rho_{\rm max}}
\newcommand \sret {s_{\rm ret}}
\newcommand \dsret {\Delta s_{\rm ret}}
\newcommand \pz {(\phi, \zeta)}
\newcommand \wn {w_n}

\newcommand \beq {\begin{equation}}
\newcommand \eeq {\end{equation}}

\def \mref#1{(\ref{#1})}

\def\la{\hbox{\hspace{1.5mm}}\raise2pt
       \vbox{\hbox{$<$}}\lower2pt
       \vbox{\moveleft9.0pt\hbox{$\sim$ }}\hbox{\hskip 0.05mm}}
\def\ga{\hbox{\hspace{1.5mm}}\raise2pt
       \vbox{\hbox{$>$}}\lower2pt
       \vbox{\moveleft9.0pt\hbox{$\sim$ }}\hbox{\hskip 0.05mm}}
\def\lasm{\hbox{\hspace{1.5mm}}\raise2pt
       \vbox{\hbox{$<$}}\lower2pt
       \vbox{\moveleft8.0pt\hbox{$\sim$ }}\hbox{\hskip 0.05mm}}
\def\gasm{\hbox{\hspace{1.5mm}}\raise2pt
       \vbox{\hbox{$>$}}\lower2pt
       \vbox{\moveleft8.0pt\hbox{$\sim$ }}\hbox{\hskip 0.05mm}}
\def\npar{\hbox{\hspace{1.0mm}}
       \vbox{\hbox{$\parallel$}}
       \vbox{\moveleft6.3pt\hbox{$\not$}}\hbox{\hskip 0.5mm}}

\slugcomment{Submitted to ApJ}

\shorttitle{Pulsar shadow}
\shortauthors{Dyks, Fr{\c a}ckowiak, S{\l}owikowska, et al.}

\begin{document}

\title{Pulsar shadow as the origin of double notches\\
in radio pulse profiles}

\author{J. Dyks\altaffilmark{1}}
\affil{Physics Department, University of Nevada Las Vegas, NV, USA}
\email{jinx@physics.unlv.edu}

\author{M. Fr{\c a}ckowiak, Agnieszka S{\l}owikowska, B. Rudak}
\affil{Nicolaus Copernicus Astronomical Center, Toru{\'n}, Poland}
\email{michalf@ncac.torun.pl, aga@ncac.torun.pl, bronek@ncac.torun.pl}

\and

\author{Bing Zhang}
\affil{Physics Department, University of Nevada Las Vegas, NV, USA}
\email{bzhang@physics.unlv.edu}

\altaffiltext{1}{On leave from Nicolaus Copernicus Astronomical Center,
Toru{\'n}, Poland}

\begin{abstract}
We present the model of eclipsing a rotating, spatially 
extended source of directional emission by a central absorber, and apply it
to the pulsar magnetosphere.
The model assumes the radially extended \emph{inward} radio emission 
along the local direction
of the magnetic field, and the pulsar as the absorber.
The geometry of the magnetic field lines of 
the rotating dipole is favourable for the double
eclipse events, which we identify with the double notches observed
in pulse profiles of nearby pulsars. For pulsars with large dipole inclinations
$70^\circ \lasm \alpha \lasm  110^\circ$
the double notches are predicted to occur within a narrow phase range
of $20^\circ - 30^\circ$ before the main radio peak. 
Application of the model to PSR B0950$+$08 establishes it as a nearly 
orthogonal rotator ($\alpha \simeq 75^\circ$, $\beta\simeq -10^\circ$)
with many pulse components naturally interpreted in terms of the \emph{inward}
radio emission from a large range of altitudes. The inward components include
the intermittently strong, leading component of the main pulse, 
which would traditionally
have been interpeted as a conal emission in the outward direction.
The model also identifies
the magnetic field lines along which the radially extended inward 
radio emission
occurs in B0950$+$08. These have a narrow range of the footprint 
parameter $s$ close to $\sim1.1$ (closed field line region, near the
last open field lines). We describe directional characteristics of
inward emission from the radially extended region and compare them
with characteristics of extended outward emission.
Our work shows that pulse profiles of at least some pulsars
may be a superposition of both inward and outward emission.
\end{abstract}

\keywords{pulsars: general --- pulsars: individual (PSR B0950+08) --- polarization --- gamma rays: theory ---
radiation mechanisms: nonthermal}


\section{Introduction}

This paper addresses two enigmatic phenomena observed in average 
pulse profiles of radio pulsars: the double notches 
of B0950$+$08, B1929$+$10, and 
J0437$-$4715 (Navarro \& Manchester 1996; 
Rankin \& Rathnasree 1997; Navarro et al.~1997, hereafter
NMS97; McLaughlin \& Rankin 2004, hereafter MR04) as well as some
puzzling profile components observed in pulse profiles, 
especially among nearby and/or bright pulsars.

The double notches have the appearance of absorption dips, or eclipse dips:
they look as slots carved in a continuous emission pattern, not as
the ubiqitous minima between different emission components
(NMS97; MR04).
The phase at which the double notches occur, their depth, 
as well as their width weakly depend on the observation frequency $\nu$ 
(NMS97; MR04).

Although known for almost ten years (Navarro \& Manchester 1996;
Rankin \& Rathnasree 1997), 
the double notches were interpreted for the first time only recently
by Wright (2004). He assumed that they
result from absorption of radio waves emitted by a region which is
extended over a very large range of altitudes, comparable with
the light cylinder radius $\rlc$. This radial extent
is obviously manifested by the fact
that the notches occur within broad emission features
extending over a large range of pulse phase. 
As demonstrated in many studies of high-energy emission
from pulsars, this effect is typical of radially extended emission region
(eg.~Morini 1983; Romani \& Yadigaroglu 1995, hereafter RY95; 
Dyks \& Rudak 2003, hereafter DR03). 
The precious idea of Wright is that the double notch effect
is produced by a \emph{single} absorber, which provides his model
with elegant simplicity. However, the location of the absorber,
and the geometry of the emitting regions are far less natural: they 
are hardly identifiable with any discernible features in pulsar 
magnetosphere. In this paper we show (Sections 2 and 3) that the single
absorber does not have to be located at significant fraction of $\rlc$
to be capable of producing the double notch effect. The pulsar itself can
do the trick, and the shape of the emission regions can be identical
to the shape of the dipolar magnetic field lines.

The pulsar can obviously play the absorber's role only when the radio emission
in the magnetosphere is \emph{inward}, ie.~is directed roughly 
``antiradially", towards the neutron star. This is what we assume in this
paper at least for the radially extended emission regions responsible
for the profile components with the notches. As we show in the accompanying
paper (Dyks et al.~2005), 
the peculiar mode changing observed in
B1822$-$09 (Fowler \& Wright 1982; Gil et al.~1994, see Fig.~4 therein) 
can be interpreted in terms of emission that
sometimes flips its direction by $180^\circ$. 

In section 3 we apply our theory of double notches to B0950$+$08.
The information we gain from the model of notches is subsequently
used to intepret enigmatic pulse profile components in this object
(Section \ref{profile}).  

Perhaps the most puzzling of all components in pulse profiles of pulsars
are the interpulses separated
from the main pulses by the phase range considerably different from 
$180^\circ$ (eg.~Manchester \& Lyne 1977; Hankins \& Fowler 1986). 
Three different geometries have been proposed to explain
this phenomenon, but no one is free from problems.
The two-pole interpretation assumes a nearly orthogonal rotator
($\alpha\sim\zeta\sim90^\circ$, where $\alpha$ is the dipole inclination,
$\zeta=\alpha+\beta$ is the viewing angle measured from the rotation axis,
and $\beta$ is the impact angle). The main radio pulse (MP)
and the interpulse (IP) are supposed to be
observed when our line of sight approaches
each of the two polar caps and the radio beams aligned with the magnetic
dipole axis come into our view. This idea is contradicted by the 
IP-MP separations
significantly different from $180^\circ$ \emph{and} by bridges of extended
emission which are often detectable \emph{only on one side of the MP}.
The other two interpretations assume a nearly aligned rotator
($\alpha\sim\zeta\sim$ a few degrees).
One of these nearly aligned interpretations (eg.~Lyne \& Manchester 1988, 
hereafter LM88) assumes that the radio beam has a form of a hollow cone
centered on the dipole axis.
The MP and the IP are observed when our line of sight
enters and exits the conal beam. This idea is contradicted by the ubiquitous,
large disproportion of intensities of MP and IP, \emph{and} by the lack
of evolution of the MP-IP separation with the observation frequency.
Another interpretation relies also on a nearly aligned
case (Gil 1983): the radio emission beam is assumed to
consist of a few nested hollow cones of different angular radius,
centered on the dipole axis. In the course of rotation our line of sight moves
mostly between two concentric hollow cones of radio emission, grazing or crossing them at/near two phases separated by 180$^\circ$. This interpretation
is contradicted by the lack of extended bridges of emission at all pulse
phases. The probability of occuring is also an issue for both of 
these nearly aligned interpretations.

B0950$+$08 is a classical example of a puzzling interpulsar.
The separation between its MP and IP is considerably different 
from $180^\circ$
and there is a bridge of radio emission connecting MP and IP.
Section 5.2 provides interpretation of the radio profile of B0950$+$08.

\section{Model of double notches}

In the next two sections we show that the idea raised by Wright (2004),
that a \emph{single} absorption region can produce \emph{double} notches
observed in the pulse profiles of three pulsars described by McLaughlin \& 
Rankin 
(2004) is \emph{generic} for the pulsar magnetosphere system.
For the inward radio emission the plasma surrounding 
the neutron star becomes a natural absorbing target and can eclipse 
the radio emission region. We show that
obscuration of a thin-walled, radially extended 
radio emission region by the pulsar itself can result in double eclipse events.
In the entire paper we assume that radio emissivity in the corotating frame
(CF) is symmetrical
with respect to the magnetic equator (as is the structure of dipolar magnetic 
field).

\subsection{Qualitative description of the double eclipse phenomenon}

\begin{figure}
\epsscale{0.8}
\plotone{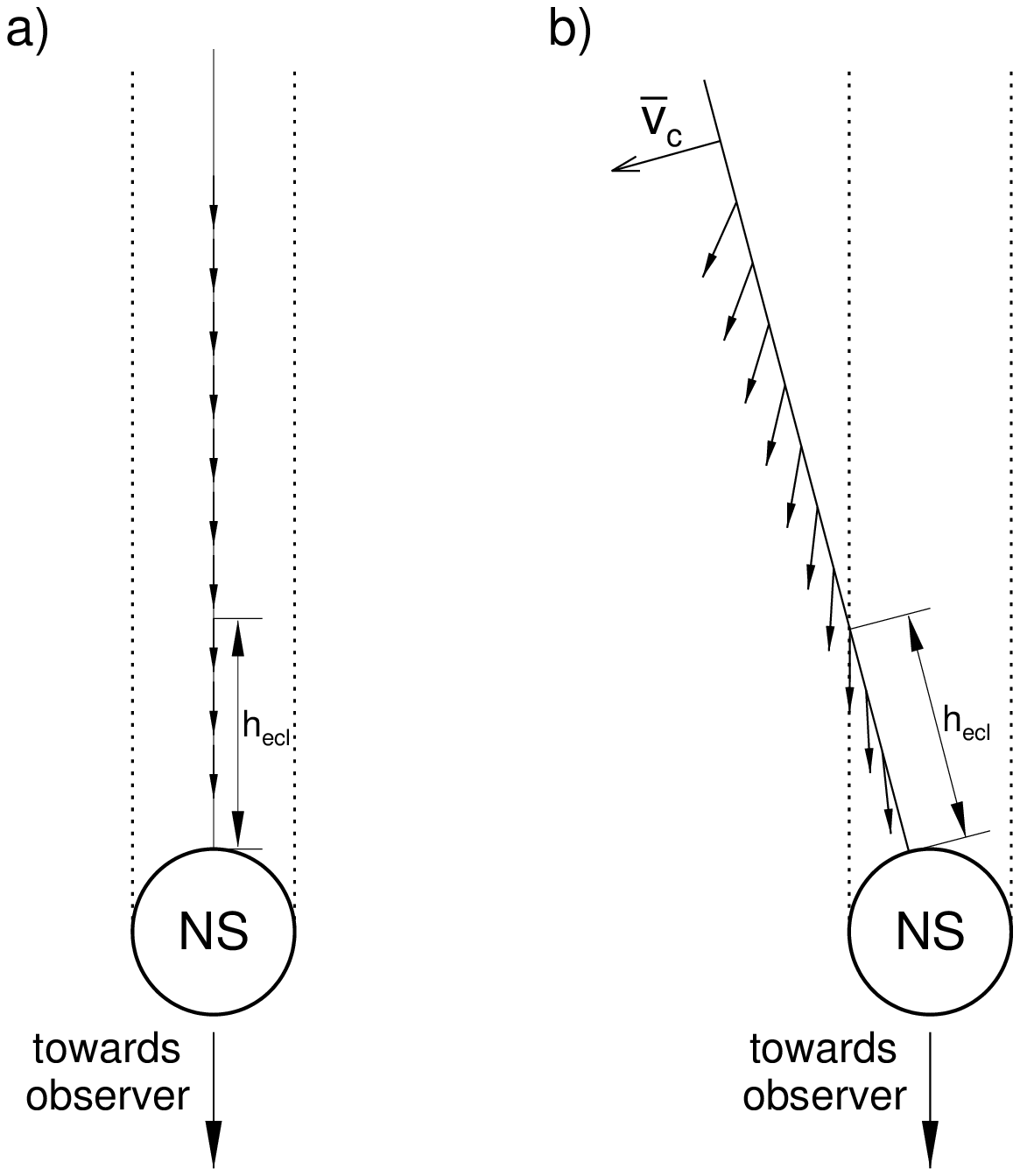}
\caption{The influence of aberration on the eclipse condition.
The plane of the rotational equator for the orthogonal rotator is shown, 
with 10 photons emitted from different altitudes
along the dipole axis in the CF (a).
Panel (b) shows emission directions of the same photons in the inertial 
observer frame (IOF). Because of the aberration effect, photons emitted
from altitudes larger than $\hecl$ are not eclipsed. General relativistic 
bending of photon trajectories is neglected in this figure.
\label{dipax}}
\end{figure}

As an introductory exercise, let us consider the situation shown in Fig 
\ref{dipax}a:
the straight dipole axis is pointing exactly away from the observer, 
and the inward emission along the axis in the CF points towards 
the observer. For simplicity, let us ignore refraction
and gravitational bending (GB) of photon trajectories.
Obviously not all radiation will be eclipsed by the pulsar.
In the inertial observer's reference frame (IOF) the emission direction
diverges from the dipole axis (Fig \ref{dipax}b) because the aberration effect
projects the emission directions ``forwards", ie.~in the direction 
inclined towards the local
corotation velocity $\vvcor = \vec \Omega \times \vec r$. 

For the central absorber of radius $\reff$, 
the radiation emitted from low altitudes below 
$\hecl = \recl-\rns \simeq (\reff \rlc)^{1/2}-\rns$, 
is absorbed/eclipsed because of small magnitude of 
the aberration effect. The radiation emitted above $\hecl$ is
not absorbed,
because it is not directed towards the pulsar in the IOF. 
Thus, for the eclipse event to occur the emission direction 
in the IOF
(ie.~the aberrated CF emission direction) must be close to antiradial.
It has to be exactly antiradial if the size of the absorber is negligible 
in comparison with the 
radial distance of the emission: $r \gg \reff$.

\begin{figure}
\epsscale{0.5}
\plotone{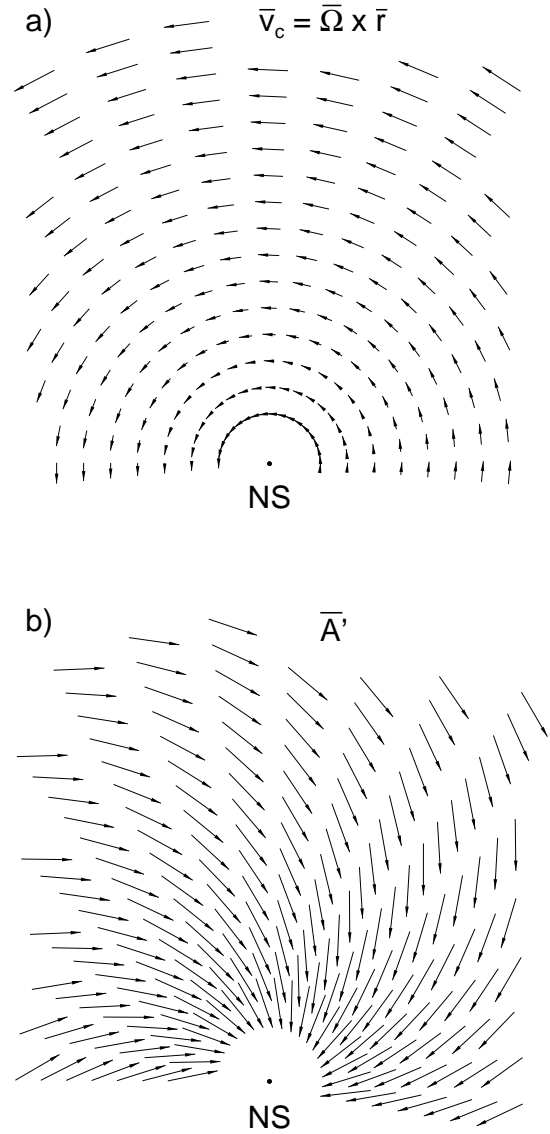}
\caption{a) Vector field of corotational velocity $\vvcor = \vomega\times 
\vrd$ in the IOF. b) Vector field of absorbed directions $\vA$
in the corotating frame
(CF).
$\vA$ is unambiguously determined by the velocity field $\vvcor$
through eq.~\mref{vA}. Photons which are emitted along the local direction of 
$\vA$ in the CF, propagate towards the center of the NS in the IOF
and are absorbed/eclipsed.
\label{fields}}
\end{figure}

Now we consider a more general situation, which will open the possibility
of the double notch effect. For simplicity
we assume that $r \gg \reff$ and we consider a two-dimensional plane
perpendicular to the rotation axis (eg.~the plane of the rotational equator).
Fig.~\ref{fields}a presents the small absorber embedded in the corotational 
velocity field $\vvcor = \vec \Omega \times \vec r$. For every point
in the magnetosphere, this velocity field uniquely
determines the ``absorbed direction", ie.~a direction of photon emission in the
CF, which becomes antiradial after being transformed to the IOF.
Thus, the velocity field in the IOF unambiguously determines 
the field of absorbed 
directions in the CF (Fig \ref{fields}b). Radiation emitted along 
the absorbed direction in the CF will be absorbed and will not reach 
the observer.  

\begin{figure}
\epsscale{0.5}
\plotone{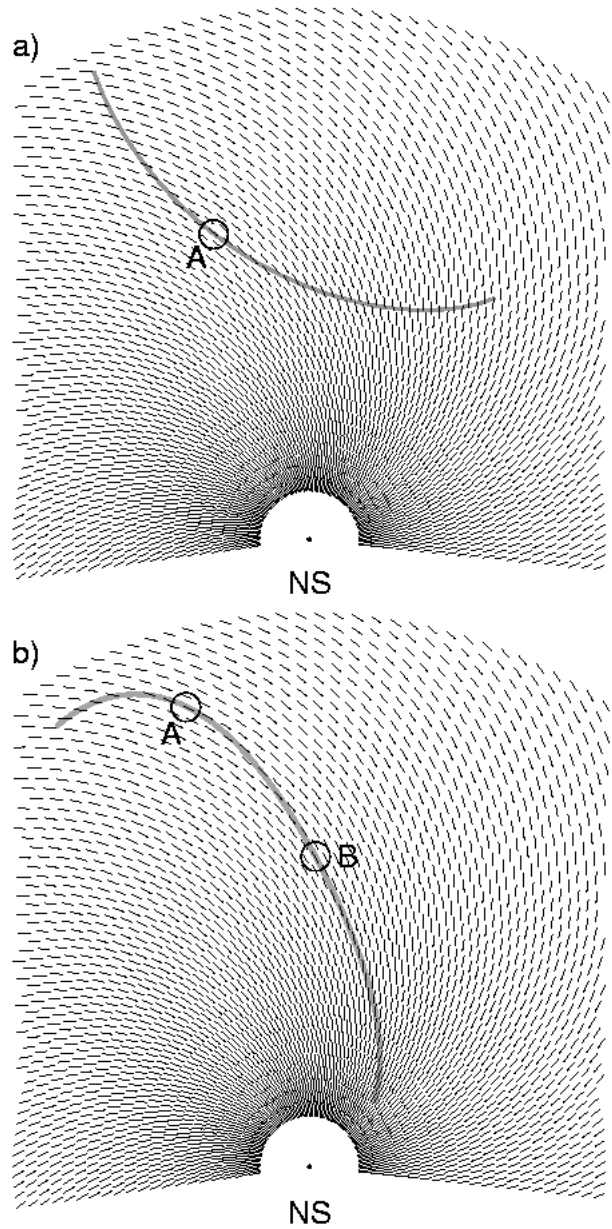}
\caption{Equatorial cross-section through an extended, thin 
emission region (grey arch) 
embedded in the vector field of the absorbed directions $\vA$ (in the CF). 
The radiation is assumed
to be inward and tangent to the emission region. a) A region tangent to
$\vA$ at the point marked with A.
The radiation emitted from A is absorbed/eclipsed, which produces
a single notch in the lightcurve. 
b) A region which is tangent to $\vA$ at two points: A, and B.
The radiation from points A and B is eclipsed. The inward radiation emitted
from the section between A and B passes on the left side of the NS,
unobscured. Inward radiation emitted from the other two segments
of the emission region passes on the right hand side. Two eclipse events
A and B occur, separated by a very short time interval.
Double notches appear in the lightcurve.
In this paper we refer to regions like A and B
with the term ``blocked regions".
\label{dublenocz}}
\end{figure}

Fig.~\ref{dublenocz}a presents a thin and radially extended
radio emission region (grey arch) embedded in the field of absorbed 
directions (both are shown in the CF). The region should be considered 
as a two-dimensional surface. Our figure only shows its cross-section 
with the equatorial plane.
We assume that in the CF 
the radio waves are emitted \emph{tangentially} to the  
emission region, in some preferred direction on the emitting surface. 
In Fig.~\ref{dublenocz} the tangent direction of emission is assumed 
to be contained within the equatorial plane.
Radiation from point A
of the region will be absorbed, because it is emitted parallel to the local 
absorbed direction. In consequence, a single notch will be observed.
Fig.~\ref{dublenocz}b presents a single extended 
emission region which will produce
double notches: the local absorbed directions
are parallel to this region in \emph{two} places: A and B.

As long as one is free to
choose \emph{any} emission region, it is hard to imagine any system of notches
which could not be interpreted in the above described way. However,
it is important to check whether the observations can
be reproduced
using an emission region which has the same morphology as the underlying
shape of magnetic field lines. Below we seek
for such regions of dipolar magnetosphere which naturally produce the double
notch effect.
In the next section we describe 
a rectilinear model of eclipsing an extended emission region by the pulsar.

\subsection{Eclipse condition}
\label{theory}

A simple condition for the ``central eclipse" of a photon 
emitted at location $\vrd$ 
and
direction $\kp$ tangent to the local magnetic field $\bfl$ in the CF,
is that the photon's emission direction in the IOF must be opposite to
radial:
\beq
\ka(\vrd) = -\rvers
\label{eclcon}
\eeq
where $\rvers = \vrd/r$,
\beq
\ka(\vrd) = \lt(\vvcor, \kp),
\label{kcf}
\eeq
\beq
\kp=\pm \hat b, \ \ \ \ \hat b = \vec B/|\vec B|, 
\label{bvers}
\eeq
and $\lt(\vvcor, \kp)$ 
represents the Lorentz transformation of the photon propagation direction
$\kp$ from the CF to the 
IOF\footnote{More exactly: 
from the Lorentz frame instantaneously comoving with 
the emission point to the IOF.}:
\beq
\ka = \frac{\kp + (\gamc +
(\gamc-1)(\vec\btcor\cdot\kp)/\btcor^2)\ \vec\btcor}
{\gamc(1 + \vec\btcor\cdot\kp)},
\label{kab}
\eeq
(DR03),
where $\gamc = (1 - \btcor^2)^{-1/2}$ and $\vec\btcor=\vvcor/c=
\vomega\times\vrd/c$.
The \emph{central} eclipse condition \mref{eclcon} is not affected 
by the GR effects 
as long as the metric around 
the neutron star (NS) has no strong azimuthal dependence 
(eg.~is close to Schwarzschild).
The values of $\vrd$ provided by condition \mref{eclcon} define the center
of the ``blocked regions" -- regions in the magnetosphere such that no inward
radiation emitted from there can reach observer because it is obstructed
by the pulsar from our sight. In Fig.~\ref{dublenocz} the
blocked regions were marked with the letters A and B.
The blocked region has the width determined by the condition:
\beq
-\ka\cdot\rvers \ge \frac{(r^2 - \reff^2)^{1/2}}{r}
\label{width}
\eeq 
where $\reff$ is the ``effective" radius of the central absorber
(the NS or the opaque plasma cocoon surrounding the NS).
In the absence of the plasma cocoon, 
refraction, and gravitational bending (GB) 
of photon trajectories, we would have $\reff = \rns$.
With the GB included $\reff > \rns$ because
on their way towards the NS, the photons'
trajectories bend towards
the axis connecting the center of the neutron star and the emission point.
If the NS is embedded in the cocoon of plasma 
(Michel 1991; Melatos 1997) which is opaque 
for the radio waves, it may well be the case that $\reff \gg \rns$
and GB is negligible. 
Eq.~\mref{width} is a more general eclipse condition than eq.~\mref{eclcon}.
It reduces to eq.~\mref{eclcon} in the limit $\reff \ll r$.

The observer located at direction $\n$ will detect
the shadows of the pulsar at the pulse phases at which the photons would
have reached him had they not been absorbed by the NS:
\beq
\phecl = -\phem + \vrd\cdot\n/\rlc + \dshap
\label{eclph}
\eeq
where $\phem$ is the azimuth of $\ka$ in the reference frame with
the observer in the $y=0$ plane, the second term is the flat spacetime
propagation time delay and $\dshap$ is the correction due to the Shapiro 
delay effect.
The minus sign at $\phem$ simply takes into account the fact that larger 
azimuths in the CF
correspond to earlier detection times.

The vector field of absorbed directions $\vA$ which we described in the 
previous section, is given by: 
\beq
\vA = \lt(-\vvcor, -\hat r).
\label{vA}
\eeq

\subsection{Failure of the small angle approximation}
\label{smallang}

In this section we use eq.~(\ref{eclcon})
to search for the double notch effect assuming the small angle 
approximation.
Wright (2004) was able to prove that a single absorber can
produce double eclipse phenomenon, in spite of the fact that he
used the small angle approximation at very high emission altitudes.
Following this simple approach\footnote{With the neutron star 
playing the role of the absorber the small angle 
theory actually becomes much simpler than
in the case considered by Wright 
(both the source and the absorber in motion).}, 
we constrain our considerations to the strictly
orthogonal rotator ($\alpha=\zeta=90^\circ$) and we aim for explaining
the double notch system of B0950$+$08. Among the three ``notched pulsars"
described by McLaughlin \& Rankin (2004), this pulsar has the largest probability 
of being the ``roughly orthogonal rotator". 
The position angle (PA) curve of this
pulsar (eg.~Fig.~4 in MR04) is typical for ``poleward viewing" (see Everett 
\& Weisberg 2001, hereafter EW01) which is improbable 
for nearly aligned pulsars.
EW01 find $\alpha \sim 75^\circ$ for this object. Becker et al.~(2004) 
interpret the X-ray pulse profile of B0950$+$08 and also conclude
that it is a roughly orthogonal rotator. 

We need to define a ``footprint parameter" $s\approx
\thsurf/\thpc$, where $\thsurf$ is the magnetic colatitude of the point at the
NS surface crossed by a considered magnetic field line.
We assume a thin-walled, radially extended emission region
with the same geometry as the surface formed by the magnetic field lines with 
some fixed $s$.
For orthogonal rotator and equatorial viewing geometry ($\al=90^\circ$, 
$\ze=90^\circ$), 
as well as in the limits of small angle
approximation (small $\rhodip/r$ where $\rhodip$ is the distance of emission point 
from the dipole axis), 
the eclipse condition \mref{eclcon} becomes:
\beq
\thb - \thm \approx \thab
\label{eclsmang}
\eeq
where $\thb$ is the angle between the CF emission direction
$\kp$ and the dipole axis
[$\thb=\arccos(\pm\vmu\cdot\kp/|\vmu|)\approx 1.5 \thm$], 
$\thm\approx s (r/\rlc)^{1/2}$ is the angle between
$\vrd$ and $\pm\vmu$, and $\thab \approx r/\rlc$ is the aberration angle
between $\ka$ and $\kp$. The equation refers to the 
leading side of the open field line region, and the ambiguity of the sign
takes into account the fact that the direction of $\vec B$ is opposite within
the two magnetic hemispheres.  
Thus, 
\beq
\frac{1}{2}\ s\left(\frac{r}{\rlc}\right)^{1/2} \approx \frac{r}{\rlc}
\label{eclsmang2}
\eeq
which for fixed $s$ has \emph{only one} solution that fulfills $r \ge \rns$.
This shows that for a radially extended emitting
surface formed by magnetic field lines with the same $\s$ (thin walled
``funnel" or ``tube" centered on $\vmu$) it is possible to observe
only a \emph{single} eclipse event, at least in the limit of small angle
approximation (ie.~for small $\thb$ or $\thm$) and for the orthogonal
geometry we considered.

\subsection{Numerical method}
\label{nummet}

It is then obvious that for highly inclined rotators, 
only the emission region located far from the dipole axis, 
for which the small-angle
approximation breaks down, has the chance to result in the  
double eclipse events (double notches).

We use the following numerical method 
to identify the blocked regions in the dipolar magnetosphere of a pulsar,
and to calculate the phases at which the notches can be observed:
1) We assume that the structure of the magnetic field can be approximated 
by the rotating vacuum dipole without the general relativistic effects.
We use eqs.~(A1) -- (A3) from Dyks \& Harding (2004) to calculate $\hat b$
and then eq.~\mref{bvers} from section \ref{theory} to get $\kp$.
2) We choose some value of $s$ and we identify all magnetic field lines
which have this $s$ (for details see section 2.2 in Dyks et al.~2004, 
hereafter DHR04).
3) We assume some value of $\reff$ 
(usually between a few $\rns$ and $\sim 30\rns$, Melatos 1997) 
and check whether the condition
\mref{width} is fulfilled anywhere along the field lines we selected
above. We probe the region of fixed $s$
which simultaneously fulfills the following 
conditions: $r > \rmin$, $\rho < \rhomax$, and $|z| < \zmax$, where $\rho$
is the distance measured from the rotation axis, and $z$ is the distance
from the plane of the rotational equator.
Typically we take $\rmin = \reff$, $\rhomax = 0.95\rlc$, and $\zmax = 5\rlc$.
When the condition \mref{width} is fulfilled, we record the spatial
coordinates $\vrd$ of such ``blocked region" and the direction of emission
$\ka$ from eq.~\mref{kab}.
4) From $\ka$ we calculate the azimuth $\phem$ of the emission direction
as well as the direction towards the observer who will detect 
this eclipse event: $\phem = {\rm atan}2(k_y, k_x)$, 
$\hat n = [n_x, n_y, n_z] = [(1 - k_z^2)^{1/2}, 0,
k_z]$. These are used in eq.~\mref{eclph} to calculate the phases at which
the observer will detect the eclipse events. $\dshap = 0$ is assumed.

\section{Numerical results}

To make clear visualization possible, we first discuss results of 
a two-dimensional calculation limited to the equatorial plane.
As argued above, this kind of calculation should be able to reproduce
the double notches of B0950$+$08, which is expected to be 
a highly inclined rotator.
To discuss numerical results for the retarded dipole 
with non-circular polar caps, we must generalize the definition of the
footprint parameter to: $\sret = \thsurf/\thpcret(\phdip)$ where
$\thpcret$ is the magnetic colatitude of the rim of polar cap measured
at the same magnetic azimuth $\phdip$ as the azimuth of the footprint point.
The definition is the same as in Yadigaroglu (1997) and Cheng et al.~(2000,
hereafter CRZ00).
For $\alpha$ considerably different from $90^\circ$, we use the open volume 
coordinate $\rovc$, for the reasons specified in DHR04.
One can usually assume that $\rovc \sim s$.

\subsection{Equatorial plane of orthogonal rotator}

\begin{figure}
\epsscale{1}
\plotone{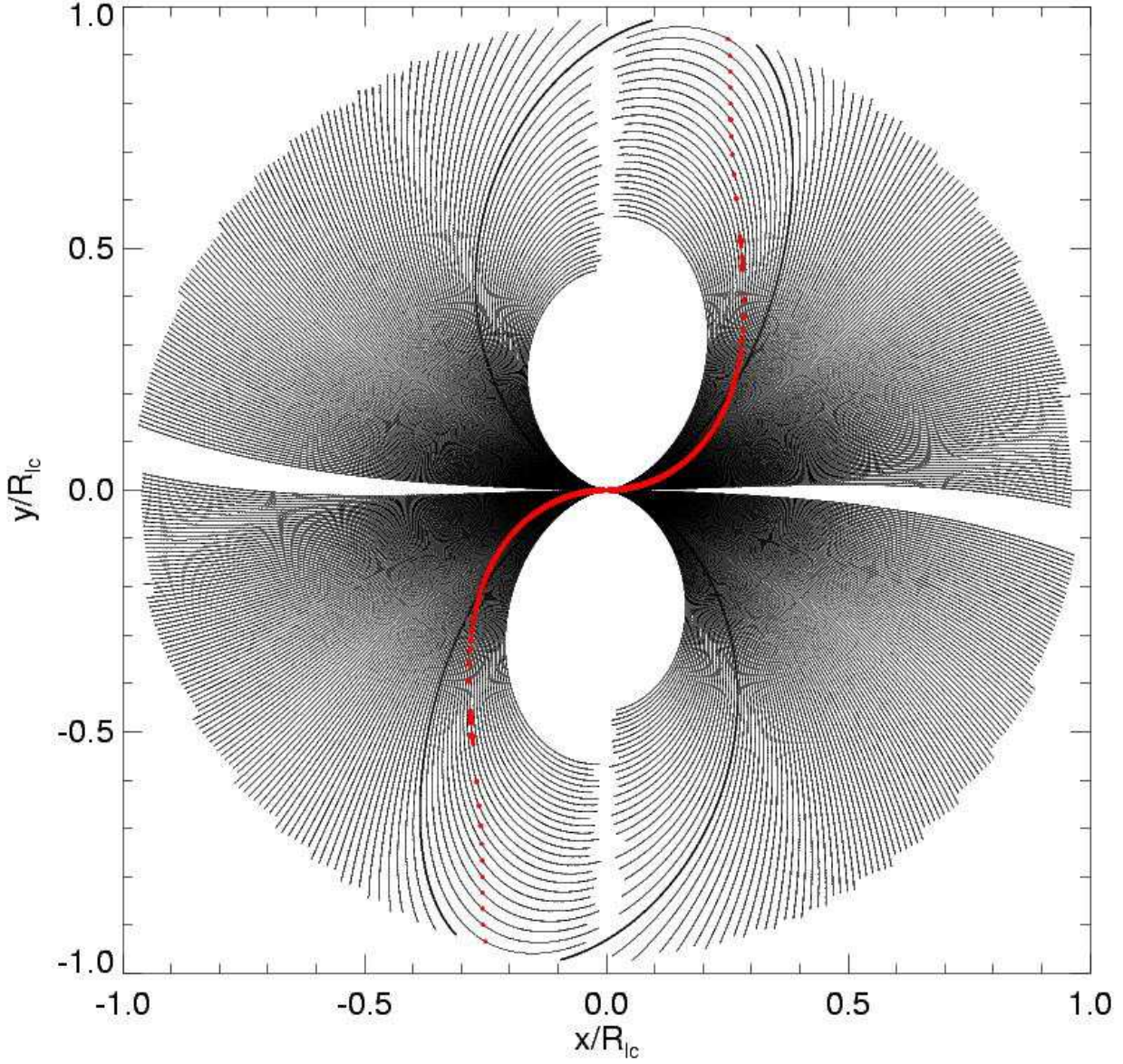}
\caption{Locations of the blocked regions on the equatorial plane
of orthogonal rotator ($\alpha = 90^\circ$). The rotation is counterclockwise.
The blocked regions
are located on the leading side of each magnetic hemisphere and
are shown with red dots. In the central parts of the magnetosphere they 
merge into S-shaped stripe. Thin lines show the magnetic field lines
of the retarded dipole. The thick solid lines are the last open magnetic 
field lines with $\sret=\rovc=1$. Two blocked regions are located
on each closed magnetic field line with $\sret$ slightly exceeding 1.
Radiation emitted from these regons is eclipsed by the NS and produces the
double notch effect. The geometry of the dipolar magnetosphere emulates
the scenario shown in Fig.~\ref{dublenocz}b. $\reff=\rns$ and the rotation 
period $P=0.25$ s was assumed in this figure.
\label{equat}}
\end{figure}

As long as our ``region of interest" is limited to the open field line region
within the light cylinder,
the application of the above-described numerical method to the
equatorial plane of orthogonal rotator gives \emph{negative} result:
open field lines with \emph{two} solutions for the eclipse condition 
\mref{width}
do not exist -- double notches cannot be produced by 
a funnel of fixed $\sret$ located within the OFLR,
at least in the orthogonal case.
However, it is sufficient to search on the closed field lines which are located
\emph{nearby} the OFLR (ie.~which have $\sret$ only slightly larger than $1$)
to find those which fulfill the eclipse condition \mref{width} at \emph{two}
different altitudes. Fig.~\ref{equat} presents selected magnetic field 
lines of the retarded dipole in the equatorial plane. Those marked with 
thicker lines are the last open field lines which have $\sret=1$. 
The numerically calculated
blocked regions are shown with red dots. 
The blocked regions associated with different field lines
form two ``blocked stripes", each of which is  
associated with a different magnetic hemisphere. 
The blocked stripes have actually a continuous form -- 
one has to connect the dots to have a correct view of them.
As expected, the blocked stripes are located on the leading side of the OFLR.
This is because in the dipolar $\vec B$ the eclipse condition can only 
hold on field lines which bend
towards the rotation direction. For lines bending backwards, the aberration
cannot make the CF emission direction antiradial in the IOF.

The magnetic field lines shown in Fig.~\ref{equat} have $\sret$ within 
the range $0.05 - 1.25$ and differ in $\sret$ by $0.01$. This allows us to 
easily identify $\sret$ for each field line, just by counting the lines 
starting from the last open ones. In this way one can find that the
field lines which are eclipsed at high radial distances ($r \gg \rns$)
have $\sret \le 1.11$. Those which are eclipsed two times
have $\sret$ between $1.0$ and $1.11$. One can identify two blocked regions
on each of these lines. For increasing $\sret$ (ie.~when one selects lines
located deeper and deeper within the closed field line region)
the two blocked regions on a given field line approach each other.
For $\sret \approx 1.11$ they merge into a single region.

For $\sret$ very close to $1$ (and $\sret > 1$) the two blocked regions
on a given field line are located very far away from each other:
one of the regions is at $r \sim 0.35\rlc$ whereas the other one 
is close to the
light cylinder. This gives a (misleading) impression that the phase separation 
between the eclipse events (between the notches) will change considerably
for emission from funnels with different $\sret$. This conjecture is wrong.

\begin{figure}
\epsscale{1}
\plotone{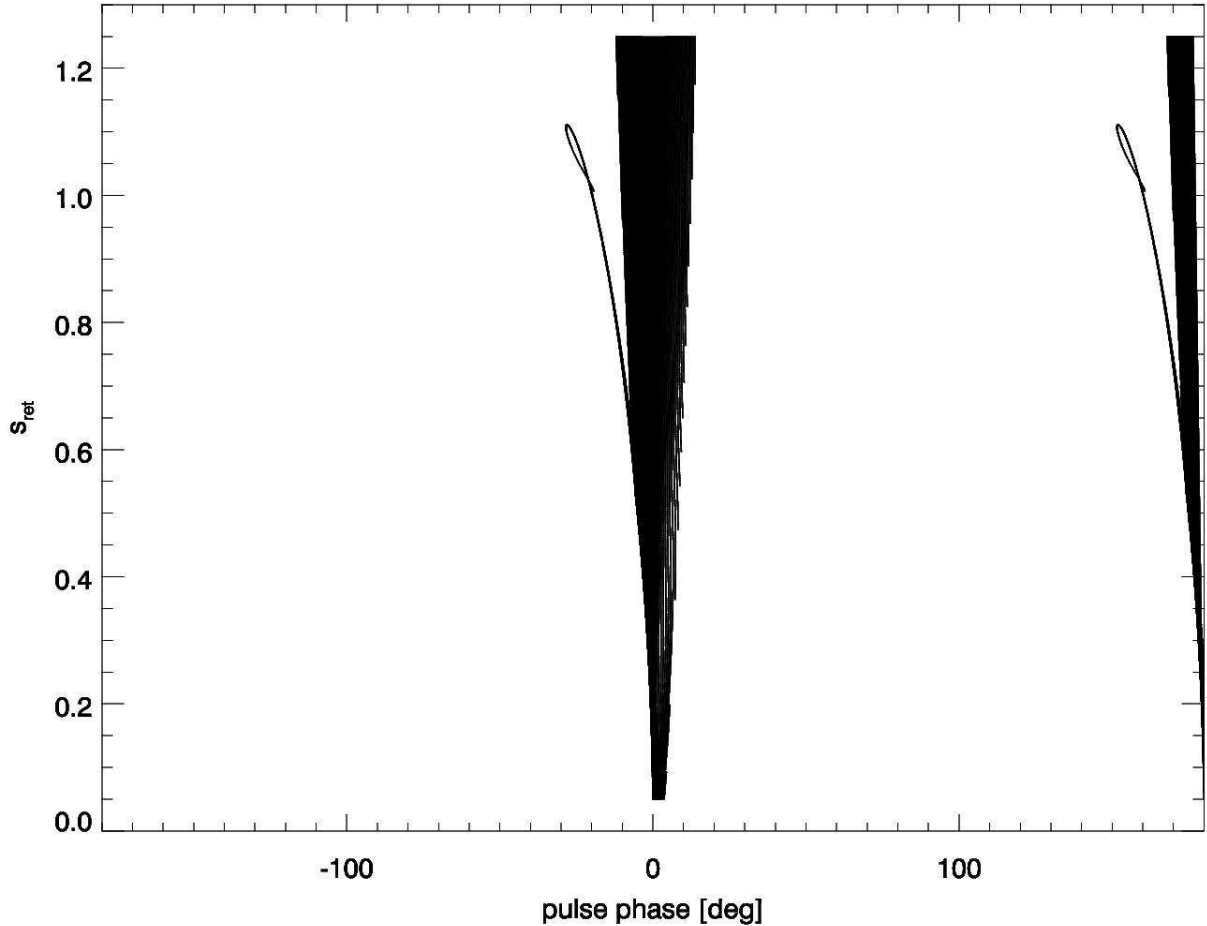}
\caption{The phases at which the eclipse events occur, for all blocked regions
shown in the previous figure (Fig.~\ref{equat}, retarded dipole). 
The footprint parameter
$\sret$ is on the vertical axis. 
Note that for emission from field lines with $\sret = 1.0 - 1.11$ the double 
eclipse occurs at $\phi \simeq -30^\circ - -20^\circ$.
The large vertical wedge at $\phi\approx 0$ represents the shadow of the NS
associated with the low altitude emission from $r \lasm 0.2\rlc$.
The loop which protrudes from its left side represents the shadow for high
altitude inward emission ($r \gasm 0.5\rlc$).
\label{ucho}}
\end{figure}

Fig.~\ref{ucho} presents the phases of eclipse events calculated with the help
of eq.~\mref{eclph} for lines with different $\sret$ (vertical axis). 
$\alpha=\zeta=90^\circ$ was assumed. The zero point of phase corresponds to
the phase of detection of
the low altitude radiation emitted along the dipole axis.
The precise definition of 
this fiducial phase is given at the beginning of section 3 of 
Dyks \& Harding (2004). 
Fig.~\ref{ucho} then tells us that for $\alpha \sim 90^\circ$
the double eclipse event (double notches) occurs at some $20^\circ-30^\circ$
before the main radio peak, \emph{just where it is actually observed for} 
B0950$+$08 (see Fig.~1 in MR04). The double eclipse can occur only
when the inward radio emission occurs at closed magnetic field lines
located very close to the OFLR, with $\sret$ between $1.0$ and $1.11$.
We do not consider this a problem because $\rlc$ is clearly an upper limit
on the size of the corotating zone (Cordes 1978).  
For the retarded dipole and for $1.00 \lasm \sret \lasm 1.04$
the two eclipses are perceived almost simultaneously, 
so that the range of $\sret$
for which \emph{easily discernible} double notches can be produced 
is $\sim 1.04 - 1.11$.
This conclusion is nevertheless sensitive to the assumed geometry of the dipolar
magnetic field.
Another important conclusion is that for $\alpha \sim \zeta \sim 90^\circ$
the double notches can occur \emph{only} within a very narrow range of phase,
which \emph{precedes} the main peak. There are no double 
solutions of the eclipse condition (eq.~\mref{width}) on the trailing side of 
the main peak. Actually, far on the trailing side of the main peak
there are no solutions for the eclipses at all (including the single ones).
The separation between the eclipse events calculated 
for a funnel of fixed $\sret$, ie.~for a horizontal line crossing 
the loop in Fig.~\ref{ucho} 
is a bit smaller than what is observed for B0950$+$08
($5^\circ-6^\circ$ versus $3^\circ$). However, it is 
sensitive to the assumed structure of $\vec B$ in the vicinity
of the light cylinder. The range of phase at which the notches occur
(leading side of the main pulse), is much less sensitive to this.

\begin{figure}
\epsscale{1}
\plotone{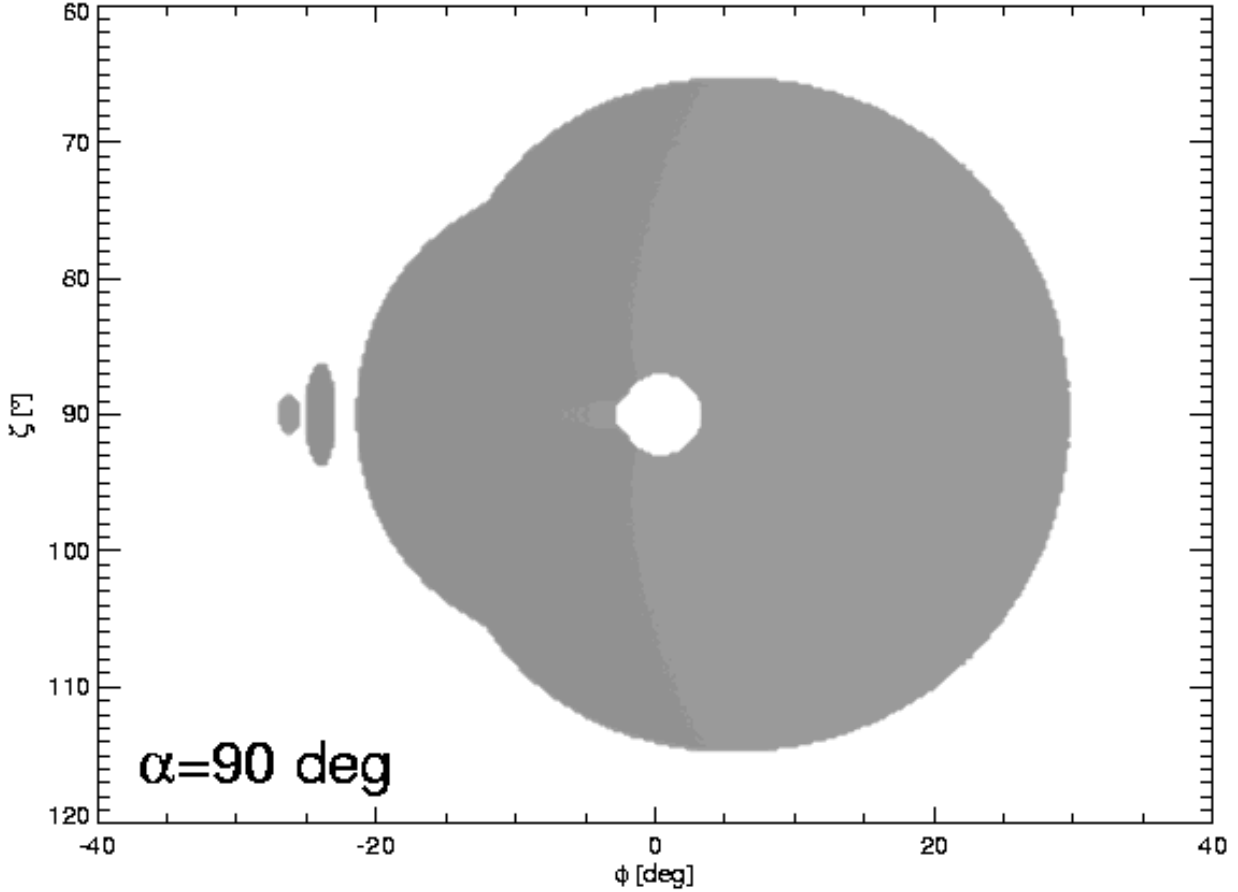}
\caption{Distribution of the \emph{eclipsed} radiation on the $\pz$ plane
for inward emission from magnetic field lines with $\sret=1.08$
and the central absorber of size $\reff = 10^7$ cm. $\alpha=90^\circ$ 
was assumed. 
$\phi$ is the rotational phase and $\zeta$ is the viewing angle.
The observer's line of sight cuts the distribution horizontally, at fixed
$\zeta$.  
The two small shadow spots
near $\phi \simeq -25^\circ$ are generated by eclipse of radiation
emitted at large $r\sim 0.5\rlc$ and are perceived as the double notches. 
The big shadow in the center is produced by the low-altitude emission.
The blank spot at its center reflects the lack of emission in the central
parts of the open field line region (no emission $=$ no shadow). 
The darkness of the pattern does \emph{not} reflect the actual flux 
of the absorbed radiation (the flux has had to be logarithmed and rescaled to fit
the viewing capabilities of the plotting program).
\label{al90}}
\end{figure}

Fig.~\ref{al90} shows the distribution of the \emph{eclipsed} radiation on 
the $(\phi, \zeta)$ plane for $\alpha =90^\circ$, $\sret = 1.08$
and $\reff = 10^7$ cm. In other words, the grey areas present 
regions on the $\pz$ 
plane within which the eclipse condition \mref{width} is fulfilled \emph{if}
the funnel of $\sret=1.08$ is assumed as the emission region. 
The spots are thus shadows of the pulsar projected
on the $(\phi, \zeta)$ plane. 
The observer's line of sight cuts the shadow spots horizontally, 
at fixed $\zeta$.
The two small shadow spots responsible for the notches
are located near $\phi=-26^\circ$ and precede in phase the large
shadow which corresponds to the low altitude emission. 
The blank spot in the center of this shadow appears because the emission is 
radiated from the funnel with $\sret=1.08$. There is no emission from the 
inner parts of the funnel, and therefore\footnote{When 
interpreting figures like Fig.~\ref{al90}
one has to keep in mind that the lack of shadow may have two reasons:
1) the emission detected at some point on the $\pz$ plane
is not obscured; 2) no emission is directed towards this point on the $\pz$ 
plane. In the case 2) we cannot see the shadow but it is not excluded
that if the radiation had been emitted towards the point $\pz$ it would
have been obscured, and the shadow would have appeared there.
The important general rule is that what is absorbed depends not only
on the characteristics of the absorber, but also on what is emitted
(ie.~on the little known characteristics of the high-altitude radio 
emission region).
A presentation of \emph{emitted} radiation on the $\pz$
plane (with shady spots here and there) does not solve this problem
because the depth of the shadows has a very different dynamical range
than the intensity range of emitted radiation. This is one of the reasons
for which it is so hard to notice the double notches for B0950$+$08. 
The double notches in its pulse profile are actually imprinted
 in a low intensity layer
of emission from high altitudes which is superimposed on a high level 
of emission of the main peak. The double notches of B1929$+$10 most probably 
are \emph{not} superimposed on 
independent emission of different origin and 
this makes them so pronounced (provided that the instrument is sensitive enough to detect this weak emission). Another viewing-related problem (other than
the different dynamical scales of shadow depths and intensities) 
is that it often happens for
a few emission layers as well as for a few shadows to overlap in pulse phase.
For example, 
the two small shadow spots tend to overlap with the large deep shadow
produced by the low altitude inward emission.}
 no shadow at $\phi=0$, 
$\zeta=90^\circ$. 
Note that Fig.~\ref{al90} was calculated for the fixed
$\sret=1.08$. Therefore, a horizontal cut through Fig.~\ref{al90} at 
$\zeta=90^\circ$ 
nearly corresponds to a horizontal cross-section
of Fig.~\ref{ucho} at $\sret=1.08$ (the correspondence would be exact
if the same value of $\reff$ had been used 
both in Fig.~\ref{al90} and \ref{ucho}).

\subsection{Non-orthogonal, highly inclined rotator}

\begin{figure}
\epsscale{1}
\plotone{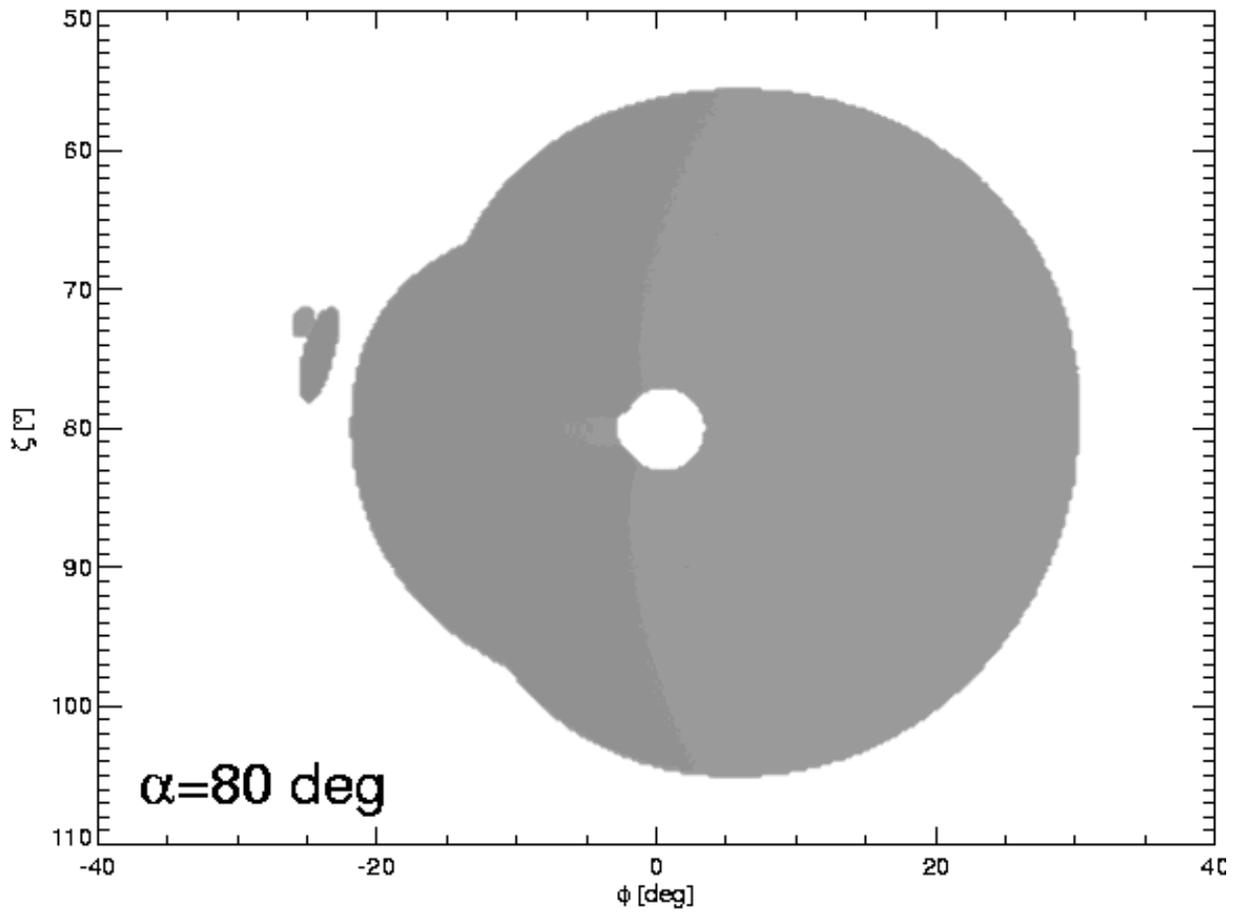}
\caption{The same as in Fig.~\ref{al90} 
but for $\alpha = 80^\circ$. Note that the notches stay pasted to each other
and both of them can be detected by a single observer.  
\label{pz80}}
\end{figure}

Because the case of $\alpha=\zeta=90^\circ$ is a very specialized one, 
it is necessary
to confirm that the double notch effect survives for $\alpha \ne 90^\circ$.
Fig.~\ref{pz80} presents the distribution of shadows
for $\alpha=80^\circ$, $\sret=1.08$, and $\reff=10^7$ cm.
Two shadow spots responsible for the double notch effect stay close to each 
other at $\phi \approx -24^\circ$. 
They are no longer at the same $\zeta$ as the magnetic pole.
They have $\zeta < \alpha$ and can only be observed
in the case of poleward viewing geometry.
This is in full agreement with the ``poleward shape" of position angle 
curve exhibited by B0950$+$08.

\subsection{The spread in $\sret$}

The results presented above were obtained for a single value of $\sret$.
The actual emission region in pulsar magnetosphere is most probably
extended across $\vec B$ and occupies a range of $\sret$.
For funnels with slightly different $\sret$, the pair of notch spots 
has slightly different location on 
the $\pz$ plane and the relative orientation of each spot in the pair
is a bit different.

\begin{figure}
\epsscale{1}
\plotone{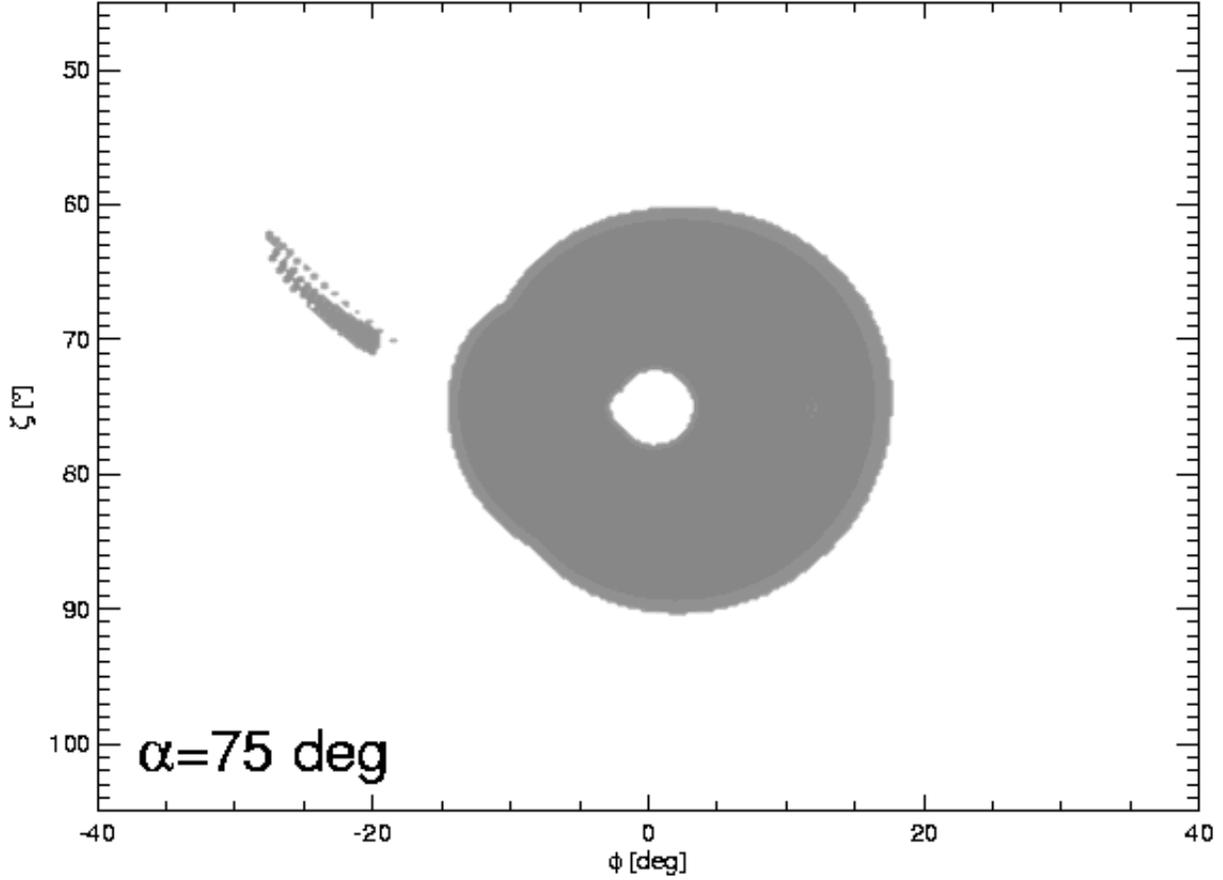}
\caption{The distribution of shadows calculated for emission covering
a range of the footprint parameter $\sret = 1.0 - 1.15$ with the step
$\dsret=0.01$, and for $\reff = 2\cdot10^6$ cm, $P=0.25$ s, 
$\alpha = 75^\circ$. 
Each set of magnetic field
lines with different $\sret$ produces different pair of the notch spots.
For different $\sret$ the pairs are located at slightly different places on the
$\pz$ plane, which creates two bands of shadow, visible near $\phi\simeq
-20^\circ$. Note, that the bands of shadow cover $\sim 10^\circ$ of $\zeta$.
This is much more than the angular diameter of the absorber
as measured from $r\sim0.5\rlc$, which is equal to $\wn \simeq 0.4^\circ$. 
\label{srange}}
\end{figure}

Fig.~\ref{srange} shows the distribution of shadows for $\alpha=75^\circ$,
$\reff = 2\cdot10^6$ cm and $\sret$ in the range $1.0 - 1.15$ with a step
of $\Delta\sret = 0.01$. Because of the spread in $\sret$ each of the notch
spots forms its own band of shadow, extending both in $\phi$ as well as in 
$\zeta$. The extension in the $\zeta$ direction considerably
increases the probability of detection of the double notches.
The two bands of shadow cover $\sim 10^\circ$ in the $\zeta$ direction, 
which is twenty times more than the angular diameter of the absorber
viewed from $\sim 0.5\rlc$. 

If the radially extended emission region is also extended in $\sret$,
then each notch in the observed pair 
is actually a superposition of shadows corresponding to different $\sret$.
Each of these notches consists then of many ``subshadows" imprinted
in emission originating from different altitude, ie.~corresponding to 
different red points in Fig.~\ref{equat}.
Therefore, the shape of the double notches (or any notches in general)
depends on the emissivity profile 
$I(r,\sret)$ along, and across the magnetic field lines, 
which complicates numerical modeling. 
The width $\wn$ and depth of the notches
may also be affected by $I(r,\sret)$. 
In the limit of the infinitely
narrow range of $\sret$, the simple-minded
method of determining the width of notches 
($\wn \simeq 2\reff/r$) reproduces the results of numerical simulations
obtained for a single values of $\sret$. 
It must be remembered, however, that the actual
shadows are most probably a convolution of emission and absorption 
from different regions of the magnetosphere.

\subsection{Moderately inclined and nearly aligned rotators}

The other two pulsars with notches -- B1929+10 and J0437-4715 probably
have smaller
dipole inclinations than B0950$+$08. The position angle
curve of B1929$+$10 implies equatorward viewing geometry, which is improbable
for $\alpha$ close to $90^\circ$. J0437$-$4715 has a pulse profile expected
for the nearly aligned rotator (NMS97).

At the present stage of developement, our numerical codes are not suitable
to identify the blocked regions responsible for the \emph{double}
notches of these objects.
It is quite easy to check, however,  whether \emph{any} shadows can appear
in the region of phase trailing the main radio pulse.
We do this by moving from point to point within the entire volume of the
light cylinder, and checking at each point whether the eclipse condition
\mref{width} is satisfied. When it is, $\phi$ and $\zeta$ are calculated
as described in section \ref{nummet}.
This method does not provide the information on $\sret$
and one cannot discern whether or not the magnetic field lines penetrating the
blocked regions can form the surfaces with geometry favourable for the
double notch effect. However, the method allows us to calculate quickly
the regions of shadow on the $\pz$ plane for different inclinations $\alpha$.

\begin{figure}
\epsscale{1}
\plotone{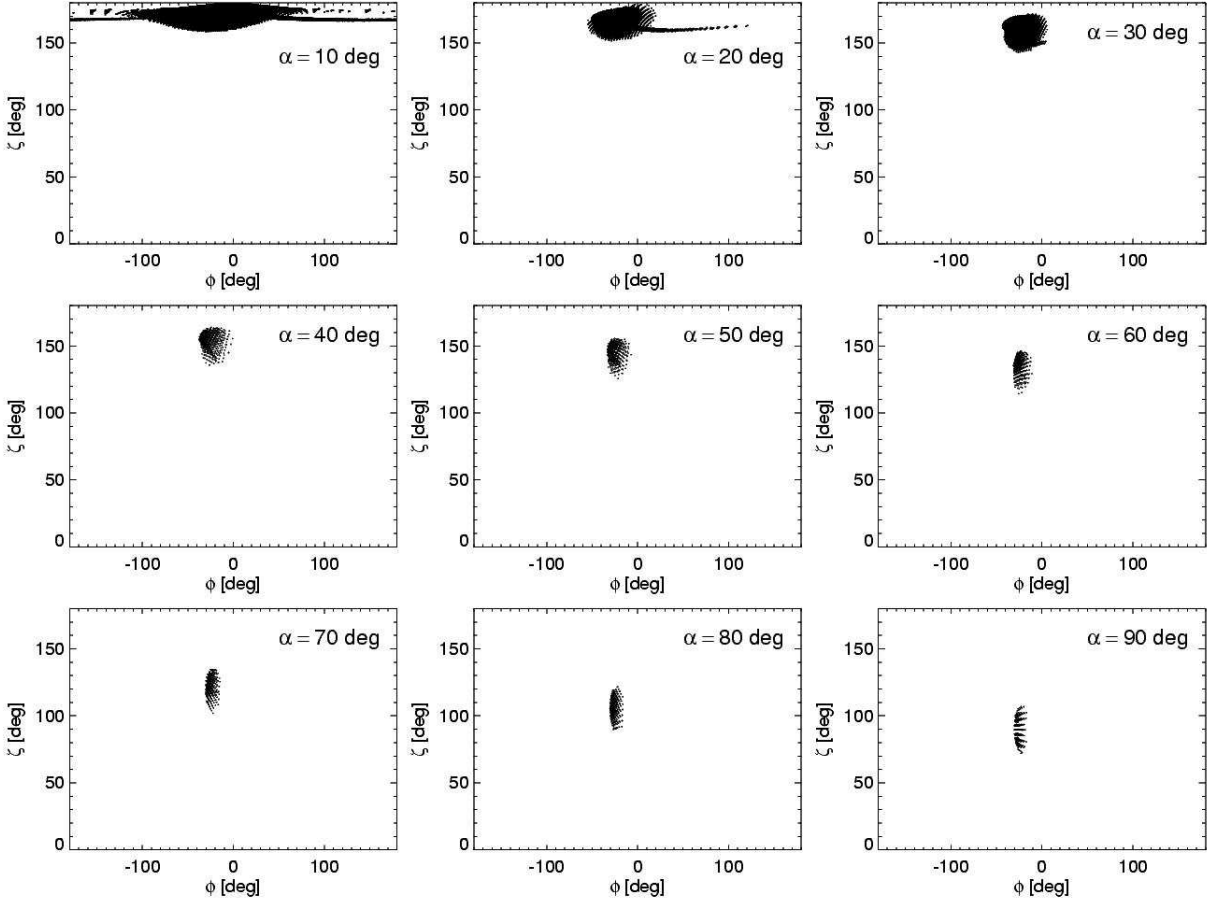}
\caption{Distribution of shadows on the $\pz$ plane for
different dipole inclinations $\alpha$ given in top right corners 
of each panel.
The figure corresponds to the emission region constrained
by the following conditions: $\rho < 0.95\rlc$, $|z| < 8\rlc$, 
$\rmin > 0.3\rlc$ 
(inward emission from nearly entire volume of the light cylinder, 
except from the small sphere around the NS). The calculations were done for
$\reff/\rlc=0.042$. The phase zero corresponds to the low altitude 
emission along the dipole axis. Note that 
for small inclination angles $\alpha \lasm 20^\circ$
the eclipse events appear far on the \emph{trailing} side of
the phase zero.
\label{cubeth}}
\end{figure}

The result is shown in Fig.~\ref{cubeth} for different values of 
$\alpha$ between $10^\circ$, and $90^\circ$, $\rmin = 0.3\rlc$,
$\rhomax = 0.95\rlc$ and $\zmax = 8\rlc$.
Inward emission from only one hemisphere is included.
One can see that for $\alpha \lasm 20^\circ$ the possibility of detection of 
pulsar shadow appears on the \emph{trailing} side of the main radio peak.
  
\begin{figure}
\epsscale{1}
\plotone{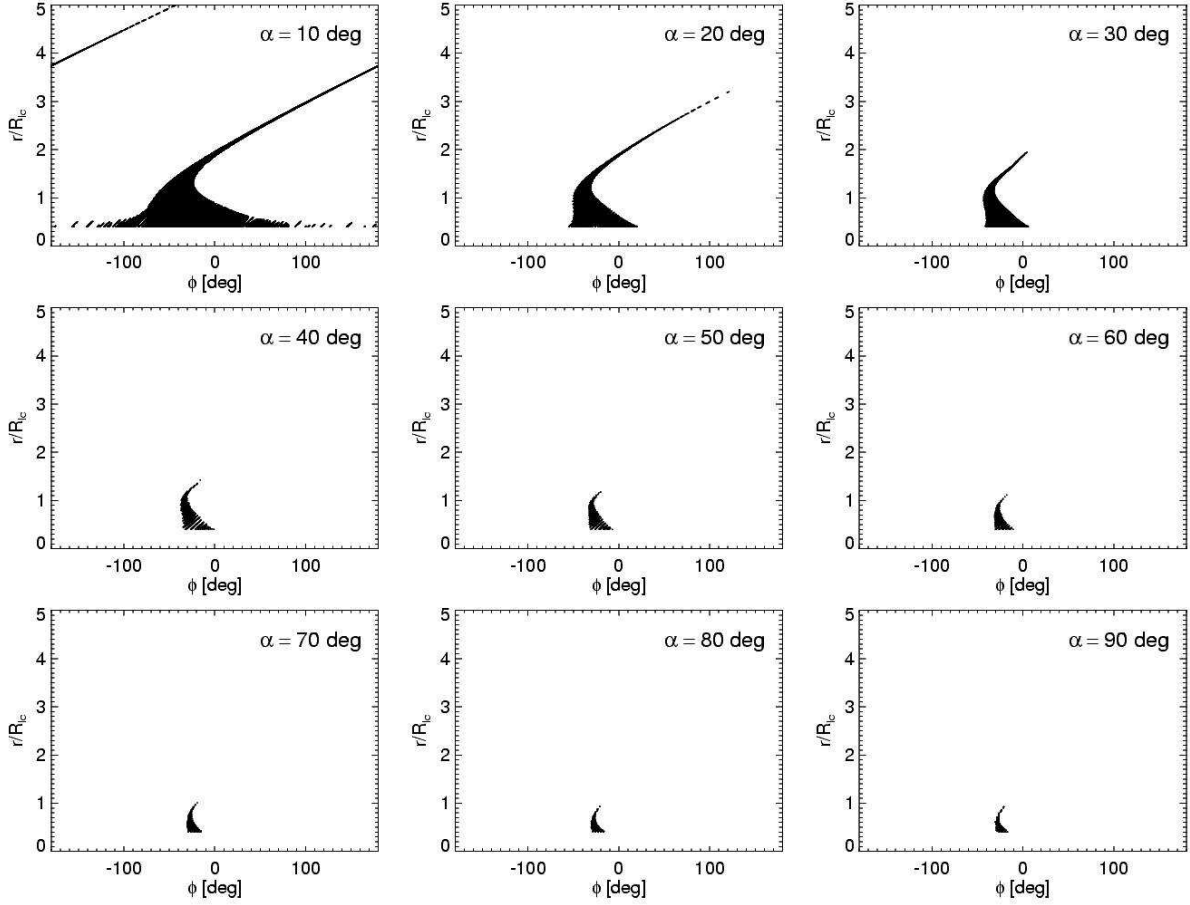}
\caption{Radial distance $r$ of blocked regions as a function of the phase
at which the eclipsed radiation emitted from them would have
reached the observer. This figure refers to the same emission region, 
and to the same blocked regions as the previous figure. It is a map of the 
points from the previous figure on the $(r,\phi)$ plane.
The eclipses far on the trailing side of the phase zero occur
because of absorption of radiation from very high radial distances $r \gasm
2\rlc$.
\label{cuber}}
\end{figure}

Fig.~\ref{cuber} presents how the same blocked regions 
map onto the space $(\phi, r)$, where $r$ is the radial distance of emission
from the blocked region. One can see that the shadows which appear on the 
trailing side of the main peak for small $\alpha$ are imprinted in emission
from very high $r > \rlc$ (but $\rho < \rlc$). The reason for which
the shadows are expected at so late phases is that the propagation time delay
(2nd term in eq.~\ref{eclph}) becomes larger for increasing $r$.

Among the two pulsars with notches on the trailing side of the main pulse,
J0437$-$4715 is consistent with the nearly aligned rotator,
with proposed $\alpha$ as small as $20^\circ$ (Gil \& Krawczyk 1997; 
cf.~NMS97). 
B1929$+$10, however, has probably quite large dipole
inclination (Rankin \& Rathnasree 1997; Wright 2004)
and poses a problem for our model.

\section{Frequency evolution of double notches}

The appearance of the double notches depends on the properties of the absorber
\emph{and} of the emission region. 
Navarro et al.~(1997) reported two types of slight frequency evolution
for J0437$-$4715: at lower frequencies the notches broaden and become more
separated.
The change of separation must result from the properties of the emission 
region, not of the central absorber.
As one can see in Fig.~\ref{ucho}, for different 
$\sret$ the separation between the notches changes slightly.
Apparently, the emissivity profile 
across the magnetic field lines has a maximum at $\sret$ which is different
for different frequencies. 
The increased width of the notches at the lower $\nu$ may be caused by
larger $\reff$. The ``radius of transparency" of a plasma cocoon
surrounding the neutron star, would be larger at smaller $\nu$,
if the plasma density decreases with $r$.
 
MR04 noted that the double notches of B0950$-$08 tend to disappear
at high frequencies (cf.~Fig.~2 and Fig.~3 therein)
or at the same $\nu$ but a different observation time.
According to our model, the double notches are imprinted in bumps
of caustically enhanced emission (see the figures and comments 
in the next section), 
which may (but do not have to) be superposed on
an underlying emission of different origin. If the spatially extended radio
emission ceases at high frequency (or at the same $\nu$ but different 
observation time), the bump in which the double notches
are immersed (and so the notches) will disappear.

The dependence of the notch width on the (frequency dependent)
radius of the central absorber is obviously worth of closer inspection 
because it
may provide invaluable information about the central object.
However, this is a very complex subject which requires exact, 3-dimensional 
modeling of the emission region. Various intensity profiles along and across 
$\vec B$ must be tested, and gravitational bending of photon trajectories 
must be included to model the cases with $\rns \lasm \reff$. 
Possibly refraction has to be included too. 
Simulations of this kind will be the subject of our future study.

\section{Pulse profile of PSR B0950$+$08}
\label{profile}

In this section we present an unorthodox, yet natural interpretation 
of the radio pulse profile of PSR B0950$+$08.
Two component radio emission region is assumed: 
in addition to a strong and narrow beam
of radio emission along the dipole axis we assume a thin, radially extended
fan/funnel of weak \emph{inward} emission from magnetic field lines 
with $\sret \simeq 1.1$. 
We assume that radio emissivity in the CF is perfectly symmetric 
with respect to the magnetic equator, i.e.~the two magnetic hemispheres
have identical emission pattern.
We associate the dipole axis beam (DAB) with the main 
radio pulse but we do not specify whether the DAB is an outward emission, 
an inward emission, or a mixture of both. 
We do \emph{not} exclude the possibility of inward emission for the DAB.
It is also possible that the funnel and the DAB are different parts of
a single, continuous emission region. In the next section we discuss
the funnel component. 

\subsection{Inward emission versus outward emission -- general remarks}

The idea of inward emission in pulsar magnetosphere is not new.
However, 
it was never studied in great detail because of a lack of firm observational 
evidence for it, and the literature on this subject is 
scarce and limited (e.g.~Cheng et al.~1986; Yadigaroglu 1997; Wright 2003).
Radially extended \emph{outward}
emission from fan or funnel-like emission regions 
is much better understood
(Morini 1983; RY95; Yadigaroglu 1997; CRZ00; DR03) and appears to be quite
successful in explaining gamma-ray profiles of pulsars.

Fig.~\ref{out} presents the pattern of \emph{outward} emission 
on $\pz$ plane, calculated for $\alpha = 110^\circ$, $\rovc=1.0$, 
$\rmax = \rlc$, and $\rhomax=0.95\rlc$. The emissivity was assumed
to be \emph{uniform} per unit length of the magnetic field lines.
One and a half of period 
is shown. The blank spots at $\pz=(0,110^\circ)$ and $(-180^\circ, 70^\circ)$
correspond to two opposite polar caps and are marked
with the letters A and B. Each of them is embedded in a
continuous emission pattern originating from the same 
magnetic hemisphere as the considered cap.
A horizontal cut through the figure at a fixed $\zeta$ produces a
lightcurve for a single observer.  
The flux received at a given phase increases with the darkness
of the pattern.

Fig.~\ref{out} illustrates that the outward emission from a large range of 
altitudes on the \emph{leading} side
of the OFLR gets \emph{spreaded} over a very large range of phases 
($\Delta\phi \sim 180^\circ$)
preceding the polar cap (and the DAB).
The outward emission from a large range of altitudes on the \emph{trailing} 
side of the OFLR gets
cumulated within a very narrow range of phase, just behind each of the
polar caps in the figure (dark arches at $\phi\simeq -160^\circ$, $20^\circ$, 
and $200^\circ$). This pile up of photons on the trailing side
of the OFLR is a well known effect of purely caustic origin, 
as described in Morini (1983)
and subsequently employed in various models of high energy emission from 
pulsars (eg.~outer gap model, RY95; two-pole caustic model, RD03).

The dotted horizontal line in Fig.~\ref{out}
cuts through the outward
emission which is detected by an observer located at $\zeta=120^\circ$.
The observer's line of sight makes the closest approach to the 
magnetic pole A
at phase $\phi=0$ where the main radio peak (MRP) is expected. 
In addition to the two caustic peaks at $\phi=40^\circ$
and $\phi=190^\circ$ the observer detects emission extended 
over a large range of phase. As emphasized by RY95, 
the brightest extended emission (usually called
the ``bridge" emission)
occurs \emph{after} the MRP, in the phase range $40^\circ - 180^\circ$.
The extended emission is weaker within the phase range \emph{preceding}
the MRP ($-170^\circ - 0^\circ$) which is called the ``off-pulse" region. 
This is a common feature of pulsar profiles observed at gamma rays 
(Kanbach et al.~1994; Thompson 2001; Grenier et al.~1988; Ulmer et al.~1995;
Kuiper et al.~2001; Fierro 1996)
which apparently are dominated by the outward emission.

Fig.~\ref{inw} presents the emission pattern from the same magnetic field lines
as in Fig.~\ref{out}, but this time the emission is inward.
The generic features of the inward emission pattern are the following:
1) The caustic pile up of photons emitted from a large range of $r$
now \emph{precedes} the main radio peak in phase and occurs on the 
\emph{leading} side of the OFLR. 2) The spreaded emission now 
occurs on the \emph{trailing} side of the MRP.
3) The bridge emission \emph{precedes} in phase the MRP. 
4) The off-pulse region follows the MRP. 

Thus, because of the change of emission direction 
most of the emission properties 
that follow from such effects as finite wave propagation speed, aberration
due to rotation -- gets reversed. For example, the relativistic
shifts of pulse components, described in Dyks et al.~(2004a) 
and Gangadhara \& Gupta (2001) occur in the opposite direction
if the inward emission is assumed.


Another feature of the inward emission map on the $\pz$ plane
is that the patterns of emission associated with different magnetic 
hemispheres switch their places in comparison with the outward case.
Thus, the polar cap marked with ``A" in Fig.~\ref{out}
is associated with inward emission pattern which surrounds ``A" 
in Fig.~\ref{inw}. Or: if the low altitude outward emission (from the vicinity 
of the NS) is located
near some $\pz$, the inward emission from exactly the same region
will be located at $(\phi-180^\circ, 180^\circ-\zeta)$. 

\subsection{Radio profile of B0950$+$08}

The interpretation of the double notches we present in Section \ref{theory}
tells us that the radially extended regions of weak radio emission
are located at field lines with $\sret \sim 1.1$.
Fig.~\ref{inw2} presents the inward emission pattern for $\alpha=70^\circ$
and $\rovc = 1.15$. By comparison with the previous figure for $\sret=1$,
we learn that many details of the pattern changed, but the generic features
discussed in the previous section (location of the caustic peaks, and of the 
bridge, with respect to the MRP) persist for $\sret \gasm 1$.  
Since the bridge observed in the pulse profile of 
B0950$+$08 \emph{precedes} the MRP we conclude
that the bridge is the inward emission. 

To identify the origin of the other components in the profile, we draw a 
horizontal line at $\zeta=120^\circ$ for the observer who can detect 
the double notches (the region in which they appear is marked with short thick
dash left of the pole B).
One can see that the notches are embedded in the caustically enhanced region
of emission, or a bump, as observed. This feature is inherent for our
mechanism of the notch generation: at low r
the radially extended emission surface
must emit radiation which passes on one side of the pulsar, 
then, at higher altitudes
on the other side, and then again on the initial side at still higher 
altitudes (these three regions of emission which passes
on both sides of the pulsar are separated by the two blocked regions;
in Fig.~\ref{equat} they can be identified with three sections
of a single magnetic field line separated by two red dots). 

One can see that the trajectory of the observer's line of sight
(the horizontal line in Fig.~\ref{inw2}) crosses two regions of emission
enhanced by the caustic effects (dark arches): 1) near $\phi\simeq-165^\circ$ 
which we identify with the interpulse, and 2) near $\phi\simeq -30^\circ$
which we identify with the notched bump described above.
Caustic enhancements of emission associated with
\emph{both} magnetic poles are located near each of those phases.
The caustic arches associated with the pole A are marked with the filled 
arrow tips, whereas those associated with the pole B are marked with the
blank arrow tips. In the case of the notched bump
emission we have no doubts that it is associated with the 
emission pattern from the pole B. A contamination from high-altitude pole-A
emission cannot be excluded, however. In the case of the interpulse the 
situation is less clear. If the radio emissivity decreases with increasing $r$,
the IP is dominated by high-altitude emission associated with the pole A.

Note that all the caustic features crossed by our line of sight
roughly overlap in phase with the difficult-to-understand components 
we observe in B0950$+$08 (the interpulse, the bridge, and the notched
bump on the leading side of the MRP).
The tendency of the modeled IP to be splitted into two or three subcomponents
is also noteworthy.

We conclude that the enigmatic components in the profile of B0950$+$08
are consistent with radially extended, weak, \emph{inward} emission from lines
with $\sret \simeq 1.1$. The shapes and locations of these components 
are mainly determined by the caustic effects. 
The pulsar is a nearly orthogonal rotator
with $\alpha \simeq 70^\circ-80^\circ$ and $\beta \simeq -10^\circ$.
The interpulse originates from very high altitudes, comparable to $\rlc$.

Hankins and Cordes (1981) noted a very interesting correlation between 
the MP and the IP: after a bright MP, a bright IP appears \emph{in the next 
pulse}. A similar phenomenon can be identified in B1055-52 (Biggs 1990).
Our geometrical model, with interpulse orginating from high altitudes, 
suggests that this effect is caused by the propagation time delay
associated with information transferred from low altitudes upwards.

\section{Polarization}

Because the projection of $\pm \vec B$ on the plane of sky does not depend
on whether we observe the inward or outward emission, we do not expect
significant modifications of the simple polarization model of
Radhakrishnan \& Cooke (1969) and Komesaroff (1970), provided
that both the inward and the outward emission represent the same mode
of radio waves.

However, 
the polarization model of Blaskiewicz et al.~(1991, hereafter BCW), which
takes into account the corotation, is affected significantly.
The BCW model can be summarized as follows: for emission from increasing
altitudes the position angle (PA) curve is shifted
 towards \emph{later} phases.
The shift, as measured with respect to the fiducial zero phase,
is equal to $\dpa \approx +2r/\rlc$ rad, as long as the emission altitudes 
are not too close to the light cylinder (Dyks et al.~2004a).

It can be shown that the corotation has the exactly opposite effect
on the PA curve if the inward emission is considered: for increasing $r$
the PA curve is shifted towards \emph{earlier} phases by $-2r/\rlc$ rad.

Our model assumes that the extended emission components/bridges are the 
inward emission which on average
originates from higher altitudes than the main peak. Thus, 
the standard S-curve fitted within the phase range which does not include 
the main pulse should reflect this anti-BCW shift toward earlier phases.
This should manifests itself in two ways:
1) The inflection point of the S-curve fitted to the high-altitude emission
should be located at earlier phase than the main pulse.
This is exactly what happens for B0950$+$08 and B1929$+$10
and is obvious in Figs.~8, 9, 10, as well as 16 and 17 in EW01. 
2) The PA data points observed under the main peak should therefore
be located on the right hand
side of the S-curve fitted to the high-altitude emission.
This effect is less pronounced but can also be discerned
in Figs.~9, 16 of EW01, as well as in Fig.~11 of Rankin \& Rathnasree
(1997).

Our value of $\beta$ is different from $\sim 20^\circ$ preferred by
EW01. We expect some discrepancy, because
according to our model, emission 
at various phases originates from different altitudes. Therefore,
application of the standard rotating vector model to different ranges of phase
can give different values of $\beta$, e.g.~EW01 find
$\beta = 13^\circ$ for the phase range including interpulse. 

The basically continuous shape of the observed PA curve
may at first glance appear disturbing, because of overlapping contributions
from two disconnected regions on opposite sides of the star.
However, for the sharp changes of PA to occur, the 
overlapping components must be polarized at significantly different angles, 
and the ratio of their linearly polarized intensities must vary very fast.
Such conditions may be, but do not have to be, fulfilled.
The difference of PA is alleviated by the fact that the projection of 
$\pm B$ on the plane of sky is the same for inward and outward emission.

\section{Inward emission within the main pulse}

We have found three independent arguments, which support the inward
direction of emission detected as the IP, the bridge, and the notched bump
in B0950$+$08: 1) the bump contains the double notches, which we interpret
as the shadows of the pulsar; 2) the geometry of the emission region implied
by the theory of the double notches naturally reproduces the locations 
of the IP, the bridge, and the notched bump; 
3) the S-swing of the high-altitude PA curve 
reveals the anti-BCW shift.

With so many components consistent with the inward emission, a natural
question emerges: is the emission of the main pulse inward too?
It is not easy to answer this question with the help of the above-mentioned
methods: Although we clearly see the trend for some kind of 
``notches" to appear
also at the maxima of the main pulses (MR04), it is hard to discard
the alternative, traditional explanation in terms of ``multiconal" outward
beam structure.
Another problem is that both the inward and the outward emission from 
\emph{low} alitudes are not affected by the relativistic phase shifts, which
are so sensitive to the flip of emission direction 
and which allowed us to interpret
the IP and the bridge. 

Therefore, we base our reasoning on intensity correlations between different 
components: Fig.~1 in MR04 reveals that the strongest part
of the main radio pulse of B0950$+$08 is splitted and
consists of two components, which can be easily discerned at 430 MHz.
The intensity of the leading component is clearly correlated
with the intensity of the IP, the bridge, and the notched bump.
This can be seen both by inspecting two different profile modes
at the same frequency of 430 MHz (cf.~Figs.~1 and 2 in MR04), as well as by
comparison of profiles at different frequencies (cf.~Figs.~1 and 3 therein).
It is obvious that the leading peak in the main pulse disappears together
with the components which we identified as inward emission.
It is natural to conclude then, that the main pulse
of B0950$+$08 includes a component produced by the \emph{inward} emission.
A look at our Fig.~\ref{inw2} reveals that the leading component of the MRP
may be either a low altitude emission from the pole B, or a high altitude
emission from one of the poles (most probably from A). A search for 
correlations between different components in a single pulse data may
solve this ambiguity. 

A similar phenomenon (weakening of conal components along with the extended
inward components at increasing frequencies) 
can easily be recognized also for J0437$-$4715 (Figs.~1 - 3 in NMS97), 
and for B1929$+$10 on the trailing side of the MP and IP 
(Figs.~5 and 6 in MR04). 

\subsection{Profile mode changes}

The trailing component of the main pulse of B0950$+$08
(the one which survives at 1475 MHz in Fig.~3 of MR01) 
may be the inward or the outward
component. If it is outward, then the immediate conclusion is that
the main pulse of pulsar profiles consists of both the outward and the inward
emission, with the latter appearing intermittently. 
Thus, we propose a time dependent ratio of inward and outward emission
as a reason for the ``long-term mode changing" visible
in Figs.~1 and 2 of MR04. 
The more common abrupt changes of profile shapes may 
well have a different origin, most probably
resulting from sudden changes of drift pattern (Wright 2003).
However, the peculiar profile mode changes of B1822$-$09 have also been 
interpreted in terms of inward emission,
as reversals of radio emission direction (Dyks et al.~2005).
  
It is worth to note that the two open field line regions above each polar 
cap, as well as the direction of $\vec B$ which permeates them, are all aligned
roughly on one axis, so that if any of them emits inward, it will contribute
to the outward emission from the other region. The neutron star, probably
surrounded by the dense plasma will be on the way, which naturally produces
the hollow cone shape of the radio beam.
The plasma may lead to birefringent refraction, possibly responsible for
the existence of the two orthogonal modes (eg.~Lyubarski \& Petrova 1998; 
Petrova 2000; Fussell \& Luo 2004).

\section{Conclusions}

Our main conclusions are:

\nin 1) We find that the geometry of dipolar magnetic field is favourable
for the double eclipse event to occur 
provided that inward emission is present in pulsar magnetosphere
in addition to canonical outward emission.
For nearly orthogonal rotators, with $\alpha \sim 70^\circ$ to
$110^\circ$,
this double notch effect occurs $20^\circ$ to $30^\circ$ before the main
radio peak. This effect may be, thus, responsible for the double notches
observed for B0950$+$08.

\nin 2) The existence of double solutions for the eclipse at the leading side
of the MRP (Fig.~\ref{ucho}) comes then naturally from the following
simple assumption:
the inward emission is
tangent to the dipolar magnetic field in the CF.
By probing the entire volume of the
light cylinder, we have only \emph{identified} the 
location of the obscured regions in the magnetosphere, and we calculated
the corresponding phases of eclipse events. 

\nin 3) The double notch effect requires that the region of inward emission
extends over a large range of altitudes and
is relatively thin. 
For the highly orthogonal rotators 
it is the thin-walled funnel formed by the magnetic field lines with
$\sret\sim 1.0$ to $1.1$. These lines are located within the closed field line 
region, but close to the last open field lines.

\nin 4) The simple geometry of emission in our model 
approximately reproduces the locations of both the absorption 
and the emission features in the pulse profile of B0950$+$08.
Based on this agreement we argue that PSR B0950$+$08 
is a nearly orthogonal 
rotator with $\alpha \simeq 70^\circ$ to
$80^\circ$ and $\beta \simeq -10^\circ$.

\nin 5) The radio emission mechanism should be capable of producing
radio emission of a given frequency over a considerable range of altitudes
($\Delta r \sim \rlc$).

\nin 6) The extended, inward radio emission from the spatially elongated
region is of very low intensity,
and can be detected only from nearby pulsars (like B0950$+$08, B1929$+$10).
For the closest pulsars, however, it is this emission which is
most probable to get into our view. It is possible to miss the strong
radio beam along the dipole axis but still detect this weak ``fan" emission.
There should be a population
of dim nearby pulsars with unusual pulse profiles. Some of them may have the
main radio peak missing and may only posses extended, bridge-like emission
components (possibly hard to detect).

\nin 7) The radial extent of the radio emission region unavoidably 
implies that the appearance of some components in pulse profiles 
is governed not
only by the spatial distribution of emissivity, but also by the caustic 
effects. We identify the interpulse of B0950$+$08 as a radio peak of caustic 
origin. The notched bump before the main peak is of the same origin.
The caustic, high-altitude nature of the IP means that B0950$+$08 is 
a kind of a ``false interpulsar", with the IP generated
in a different way than those described in Introduction. 
Fig.~\ref{inw2} demonstrates that the probability
of observing such ``caustic interpulsar" is very large: most horizontal 
cuts
through the figure correspond to lightcurves with interpulses. 
However, because the fan/funnel component of the radio emission is weak,
the caustic interpulses should be detectable only from the nearest pulsars.
Interestingly, most known interpulsars are close objects (Manchester
et al.~2005).

\nin 8) We interpret
 the enigmatic components in the pulse profile 
of B0950$+$08 as inward components. The strong, leading component 
in the main pulse correlates
with them and apparently is also the inward emission.
This interpretation implies that the long-term 
profile mode changes observed for B0950$+$08 are generated 
by time-dependent contributions of inward emission to the outward beam. 

\nin 9) With the bridges of high-altitude, low-intensity emission 
connecting smoothly
with the main radio peaks, the geometrical method of determining the radio
emission altitude breaks down, especially if the pulse width is measured
at the ``lowest detectable" intensity level (Kijak \& Gil 2003). 
Given that the radio emission region for a given frequency extends
over a large range of altitudes and at least for some objects
has the form of the fan beam, the principle underlying the method
needs to be reconsidered. The pulse width measured 
at the lowest detectable emission level may at best correspond
to the radial distance from which this \emph{edge/wing} emission is measured
(if the actual $s$ is close to 1). It may have little, or nothing to do
with the altitude from which the bulk of emission in the main pulse
originates.

\nin 10) For the same reason the delay-radius method of determining $r$ breaks
down.
Not only the emission does not originate from a fixed altitude, as 
assumed in BCW, but the pulse profiles posses inherent asymmetry caused by 
the contributions from high-altitude inward components (eg.~the notched bump
in B0950$-$08 broadens the left wing of the main peak and destroys
any low altitude symmetry which the main peak might otherwise posses). 

\nin 11) The strongly asymmetric pulse profile of B0950$+$08
is produced by the purely symmetric emission region which follows
the symmetry of the dipolar magnetic field. No ``higher mutipoles" or 
``sunspot-like" structures (eg.~Gil et al.~2002) 
are required to understand it.
It is the rotation which naturally breaks the symmetry.

\nin 12) The locations of components, bridge and double notches
in the pulse profile of
B0950$+$08, as well as the occurence of the anti-BCW
shift in B0950$+$08 and B1929$+$10, all can be interpreted in terms of
the \emph{inward} radio emission. The two-directional emission
can also be inferred from the peculiar mode changes of B1822$-$09
(Dyks et al.~2005). 
Apparently, some radio data are less difficult
to interpret if one admits inward emission in pulsar magnetosphere.

\section{Discussion}

The inclusion of the inward radiation into the radiation generation
scheme opens new possibilities that can help explaining the
variety of pulsar radio profiles (Rankin 1983;
Lyne \& Manchester 1988; Kramer et al.~1998; 
Xilouris et al.~1998; Gould \& Lyne 1998; Weisberg et al.~1999).
Among others,
refraction and reflection of radio waves in plasma are the processes
that we are looking forward to analyze in greater detail. As already
showed by Petrova (2000) the observed core component can be the result
of refracting of part of the conal emission. Some part
of the inward emission that is eclipsed by the star can undergo heavy
physical processes in dense plasma and possibly manifest itself to the
observer.

The peculiar mode changing of B1822$-$09 suggests that 
\emph{reversals} of the radio emission direction are possible.
Periodical changes to the emission direction can lead (under certain
conditions) to subpulse drifting or to pulse nulling or
odd-even mode change. 
This can indeed be a competitive model for a ``carousel of
sparks". 
The model could possibly explain such exotic phenomena as the 
bi-drifting observed for J0815$+$09 (McLaughlin et al.~2004)
as well as the intermittent nature of 
the pulsars J1649$+$2533 and J1752$+$2359 (Lewandowski et al.~2004).
However, a detailed
time-dependent model of the pulsar magnetosphere is crucial to explore
physically justified scenarios.

\begin{figure}
\epsscale{1}
\plotone{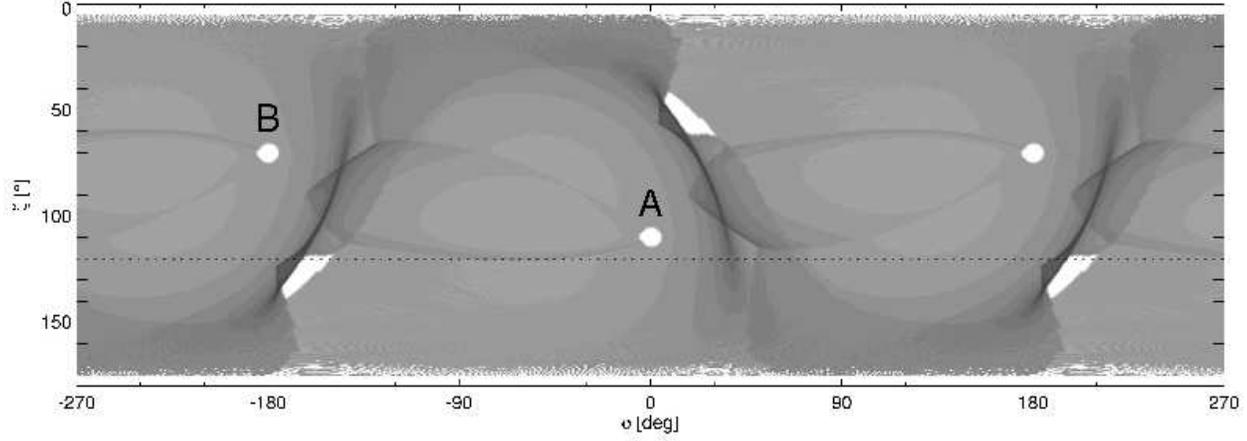}
\caption{Distribution of \emph{outward} radiation emitted along the last
open field lines in the CF on the $\pz$ plane for $\alpha=70^\circ$.
Two opposite polar caps are marked with letters A and B. 
Each one is followed by a dark arch of a caustic enhancement of emission,
which occurs on the trailing side of the open field line region.
The observer located at $\zeta=120^\circ$ views this pattern along the dotted
horizontal line. He successively records the main radio pulse at the closest 
approach
to the pole A (near $\phi=0$), then the caustic peak near $\phi=40^\circ$
followed by the bridge emission within $40^\circ \lasm \phi \lasm 190^\circ$
and finally another caustic peak near $\phi=195^\circ$. 
The figure was calculated
for $\rhomax = 0.95\rlc$ and $\rmax = \rlc$.
\label{out}}
\end{figure}

\begin{figure}
\epsscale{1}
\plotone{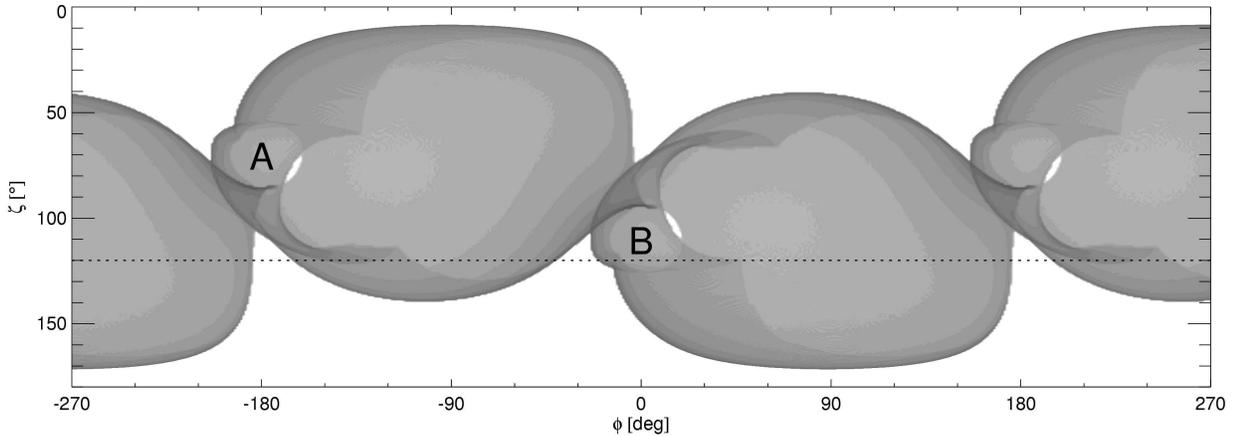}
\caption{Distribution of radiation from the region with the 
same geometry
as in the previous figure but for the \emph{inward} emission.
Note that the two magnetic poles switched their positions, and that
the main features of the outward emission pattern (the caustic peaks
and the bridge emission) jumped to the leading side of the phase zero. 
\label{inw}}
\end{figure}

\begin{figure}
\epsscale{1}
\plotone{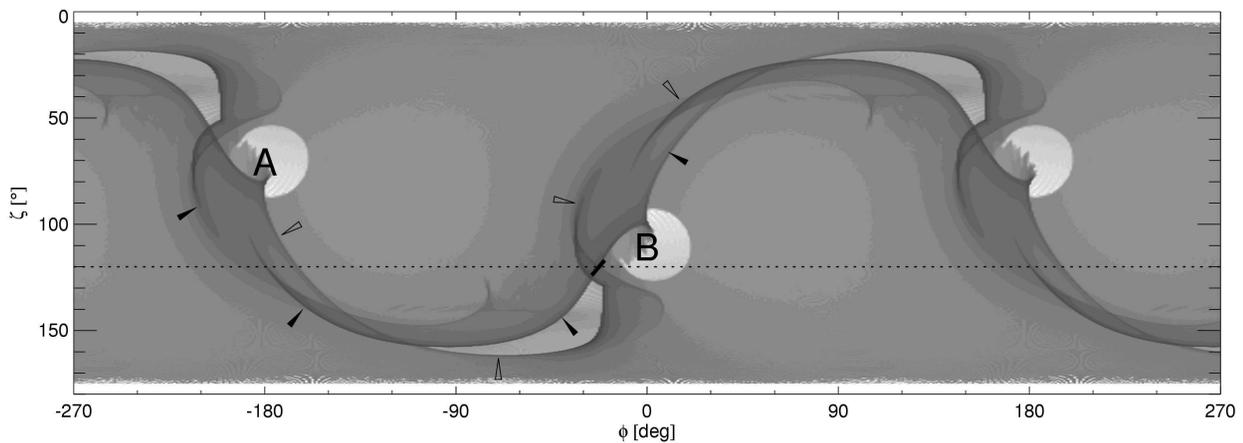}
\caption{Distribution of the \emph{inward} emission on the $\pz$ plane
for the same $\alpha$, $\rhomax$, and $\rmax$ as before, but for a slightly 
different footprint parameter $\rovc = 1.15\sim\sret$.
The filled arrow tips mark the emission pattern of the pole A, whereas the open
tips mark the pattern of pole B. Short thick dash left of B marks the region
where the double notches occur. The line of sight of 
observer who detects the notches, also crosses two regions of caustically
enhanced emission, near $\phi \sim -165^\circ$ and $-30^\circ$, as well as 
the bridge of emission between them. The same arrangement of components
is observed in the radio profile of B0950$+$08.
\label{inw2}}
\end{figure}

The gamma-ray profiles of pulsars are most naturally interpreted
in terms of \emph{outward} emission from fan/funnel-shaped regions
extending over large range of altitudes (eg.~RY95; DR03), although
the contribution of inward radiation is not excluded.
The X-ray, and UV-emission is much less understood.
B0950$+$08 is an X-ray pulsar. In spite of similar pulse profiles
in X-rays and radio (Becker et al.~2004) the main X-ray peak (the strongest
peak in Fig.~7 of Becker et al.~2004) lags the MRP
by some $30^\circ$ (Becker et al.~2004; Zavin \& Pavlov 2004).
This makes the interpretation of the X-ray profile of this pulsar unclear.
The main X-ray peak could be interpreted in terms of the
two-pole caustic model [TPC], as the caustic peak for 
outward emission, which forms on the trailing side of the OFLR
(Fig.~\ref{out}). The outer gap model with no emission from 
below the null charge 
surface (NCS) cannot explain the X-ray profile of B0950$+$08, although
recent revisions of this model do allow for the emission below NCS
(Hirotani et al.~2003). 
The broad X-ray peak which precedes in phase
the MRP can only be interpreted as the outward TPC emission if
the outward X-ray emissivity has a maxium at relatively low altitudes 
($r \sim 0.3\rlc$). The emissivity has to cease above $r\sim 0.5\rlc$,
to avoid producing the trailing peak of the TPC model, which
should roughly overlap in phase
with the IP (cf.~Figs.~\ref{out} and \ref{inw2}).
A similar disappearance/fading of 
the trailing peak of the TPC model (expected at some $160^\circ$
before the MRP)
seems to occur also in the X-ray profile of the Vela pulsar (Harding et 
al.~2002), which has similar emission geometry as B0950$+$08 
(Radhakrishnan \& Deshpande 2000; DHR04).
Fig.~1 of Harding et al.~(2002) demonstrates that the trailing gamma-ray 
peak becomes much 
less visible at X-rays. Not so for the leading peak which remains prominent
within the entire energy range of RXTE.
Interestingly, the new soft features visible below $\sim 10$ keV
(their ``Soft Pk 2" and ``Pk 3") emerge at phases where the caustic peaks
for inward emission occur (see Fig.~\ref{inw2}). 

The subject of giant radio pulses (eg.~Johnston \& Romani 2003; 
Cusumano et al.~2003) is closely related to the problem of unification of
radio with high-energies. It becomes more and more widely accepted
that the peaks in gamma-ray pulse profiles mostly have the caustic origin 
(Morini 1983; RY95; CRZ00; DR03; DHR04) 
and they appear to be dominated by the outward emission. 
It seems natural to propose the same (ie.~the caustic enhancement of the 
outward emission) as the origin of the giant radio pulses, at least
for those which are detected in phase with gamma-ray peaks.
Fixing the direction of X-ray emission (inward/outward), 
would shed light on the origin of the X-ray-coincident giant radio pulses.
 
The geometry of the inward emission region which we inferred from the model of 
notches of B0950$+$08 resembles the geometry of TPC model of DR03
as well as the geometry of the outer critical cone identified by Wright (2003).
Among physical models, the slot gap model of Arons (1983) and Muslimov \& 
Harding (2003) has most similar geometry, but for large inclination angles 
$\alpha$ it suffers from the problem 
of ``unfavourable" magnetic field lines, even when the general relativistic
inertial frame dragging is included. Interestingly, our geometrical model
of inward radio emission in B0950$+$08 unambiguously points at the high
inclination and at the unfavourable magnetic field lines. 

\acknowledgments

JD thanks Geoff Wright for the discussions we had at GSFC a year ago.
This work was supported by a research grant at UNLV
and NASA NNG04GD51G (JD and BZ), by 2P03D.004.24 (BR), 
and PBZ-KBN-054/P03/2001 (MF and AS).

\onecolumn

\clearpage

\end{document}